\documentclass[a4paper]{article}

\usepackage[english]{babel}

\usepackage{amsmath,amssymb,multirow}

\usepackage{amsthm}
\usepackage{mathtools} 
\usepackage{graphicx}
\usepackage{psfrag}
\usepackage{cuted}
\usepackage{textcomp}

\newtheorem{defi}{Definition} 
\newtheorem{theorem}[defi]{Theorem}
\newtheorem{prop}[defi]{Proposition}
\newtheorem{lemma}[defi]{Lemma}
\newtheorem{remark}[defi]{Remark}

\newtheorem{ass}[defi]{Assumption}
\newtheorem{problem}[defi]{Problem}

\newtheorem{alg}[defi]{Algorithm}
\newtheorem{notat}[defi]{Notation}

\usepackage[colorlinks=true, allcolors=blue]{hyperref}

\newcommand{\R}{\mathbb{R}}
\newcommand{\C}{\mathbb{C}}

\DeclareMathOperator{\rk}{rank}
\DeclareMathOperator{\im}{im}

\DeclareMathOperator{\Span}{span}

\def\sgn{\mbox{sgn}}
\def\real{\mathbb{R}}
\def\nat{\mathbb{N}}
\def\complex{\mathbb{C}}
\def\ra{\rightarrow}

\def\bmat{\left[ \begin{array}}
	\def\emat{\end{array} \right]}

\def\gR{{\mathcal R}}

\def\gV{{\mathcal V}}

\def\bn{\begin{enumerate}}
	\def\en{\end{enumerate}}
\def\bq{\begin{eqnarray}}
	\def\eq{\end{eqnarray}}
\def\bqn{\begin{eqnarray}}
	\def\eqn{\end{eqnarray}}
\def\bqta{\begin{eqtarraya}}
	\def\eqta{\end{eqtarraya}}
\def\bqtb{\begin{eqtarrayb}}
	\def\eqtb{\end{eqtarrayb}}
\def\bqtc{\begin{eqtarrayc}}
	\def\eqtc{\end{eqtarrayc}}
\def\be{\begin{equation}}
	\def\ee{\end{equation}}

\title{Non-overshooting output shaping for switched linear systems under arbitrary switching using eigenstructure assignment}

\author{K. Wulff\thanks{Control Systems Group, Technische Universt\"at Ilmenau,  P.O.~Box 10 05 65, D-98684 Ilmenau, Germany. \texttt{kai.wulff@tu-ilmenau.de} (Corresponding author)} \and 
	M. C. Honecker\thanks{Control Engineering Group,  Technische Universt\"at Ilmenau, P.O.~Box 10 05 65, D-98684 Ilmenau, Germany.} \and
	R. Schmid\thanks{Dept. of Electrical and Electronic Engineering, University of Melbourne, Grattan Street, Parkville, Victoria, 3010, Australia} \and 
	J. Reger\thanks{Control Engineering Group,  Technische Universt\"at  Ilmenau, P.O.~Box 10 05 65, D-98684 Ilmenau, Germany.}
}

\begin{document}

\maketitle

\begin{abstract}
We consider the analytical control design for a pair of switched linear multiple-input multiple-output (MIMO) systems that are subject to arbitrary switching signals.
A  state feedback controller design method is  proposed to obtain an eigenstructure assignment that ensures that the closed-loop switched system is globally asymptotically stable, and the outputs achieve the non-overshooting tracking of a step reference.
Our analysis indicates whether non-overshooting or even monotonic tracking is achievable for the given system and considered outputs and provides a choice of possible eigenstructures to be assigned to the constituent subsystems.
We derive a structural condition that verifies the feasibility of the chosen assignment.
A constructive algorithm to obtain suitable feedback matrices  is  provided, 
and the method is illustrated with numerical examples.
\end{abstract}

\section{Introduction}
Switched systems find application in  numerous engineering problems  such as congestion control \cite{ShortenKWL2007,ChenJX2023}, 
automotive \cite{SolShoWulOCa08,VargasCIB2022}, traffic control \cite{RastgoftarLJ2023}, fault tolerant control \cite{ZhaiLLZ2016}, robotics \cite{BabiarzCK2014}, regenerative power systems \cite{PalejiyaHMC2013,LiLLH2015,PalejiyaC2016},  aircraft control systems \cite{WangZS2016,LianLX2017,ZhaoZFSC2020,ShiS2021}, precision motion systems \cite{HeertjesN2012}, electro-hydraulic systems \cite{YuanDSB2017,XieLYY2019} and many more.
Accordingly switched systems have been studied for more than two decades by now resulting in several monographs on this subject \cite{Lib03,SunGe05,SunGe11}.
Presumably the majority of contributions is dedicated to the stability analysis of this system class, see \cite{LibMor99,DecBraPetLen00,ShoWirMasWulKin07,LinAnt09} and references therein.
With an increasing number of stability results available the focus moved towards control design methods.
Many proposed design approaches employ numerical tools like LMI methods, $H_\infty$ or set-theory approaches \cite{SunP2010,FiacchiniJ2014,FiacchiniGJ2016,FiacchiniT2017,YangZS2022}.
Model-based control approaches have been investigated, e.g. in \cite{YuanDSB2017,ZhangZB2016}, where the former employs a model reference adaptive control approach, and the latter considers a switched model-predictive control approach.

In this paper we consider the set-point tracking design problem for switched systems under arbitrary switching using eigenstructure assignment.
Many studies on the tracking problem for switched systems utilise the switching sequence of the system.
In \cite{ZhangZ2021} a state-dependent switching signal is designed to achieve non-overshooting set-point tracking.
In \cite{Sun2018} a computational procedure is proposed to obtain a switching mechanism that stabilises the input-free system while guaranteeing an upper bound for the output-overshoot.
\cite{BagliettoBT2013} considers the tracking problem for switched systems with an unknown switching sequence. 
A mode-estimator is used to apply a suitable controller into the control loop. 
It is shown that exponential stability can be guaranteed for any slow-on-the-average switching sequence even if the plant mode is not always available.
\cite{YangLQ2019} considers output tracking with dynamic output feedback for delayed switched linear systems where the switching signal is used as additional control. 
They propose a state-dependent switching rule to achieve their goal.

Eigenstructure assignment methods have been applied to switched systems subject to arbitrary and unknown switching in order to design switched controllers that achieve stabilisation
\cite{WulWirSho09,HaiBra13,Bajcinca.2015,Hai16,WulHR21,HonGWTR2023}.
By resorting to classical linear control methods these approaches provide design procedures that are familiar to control engineers.
In \cite{Bajcinca.2015} the considered subsystems share a common right invariant eigenspace. 
The devised controller assigns the left eigenspace of the closed-loop subsystems to ensure stability for arbitrary switching. 
This can also be achieved by simultaneous triangularisation of the subsystems \cite{HaiBra13,Hai16}, where generic sufficient conditions are investigated using Lie-algebra methods.
\cite{HonGWTR2023} provides a parameterisation of feasible eigenstructure assignments that guarantees stability for arbitrary switching and allows for further refinement to achieve desired transient performance.

Our approach in this paper adopts design methods for set-point tracking proposed for LTI systems.
The control design problem to achieve globally  monotonic tracking of step references for linear time-invariant (LTI) multiple-input multiple-output (MIMO) systems  is investigated in \cite{SchN2010, NtoTSF2016,XavierBS2018}.
While \cite{NtoTSF2016} focuses on the controller concept for multiple-input multiple-output (MIMO) linear time-invariant (LTI) systems to achieve monotonic tracking of a step reference at the output, \cite{SchN2010} deals with specifying a non-overshooting output for a step reference for MIMO LTI systems. The objective here is to ensure that the output error does not change sign.
By combining the stability result for switched linear MIMO systems in \cite{HonGWTR2023} with the approach on monotonic tracking for LTI MIMO systems in \cite{NtoTSF2016}, the design procedure proposed in \cite{HonSWR24} achieves stability for arbitrary switching while also ensuring monotonic outputs for step references.
A significant limitation of the algorithm proposed in \cite{HonSWR24} is that it does not provide any condition on the system structure to ensure the solvability of the monotonic tracking problem.

This contribution extends these results to obtain a generalised control concept for shaping the outputs for a broader class of switched linear MIMO systems.
Therefore we combine the concepts from \cite{HonGWTR2023} and \cite{SchN2010} to achieve globally stabilising state-feedback for switched linear MIMO systems while ensuring a non-overshooting step reference tracking.
We propose a control design method to obtain an eigenstructure assignment for the closed-loop system that ensures global asymptotic stability for arbitrary switching and achieves various design goals for the step-reference tracking of the outputs.
In particular we employ a switched state-feedback controller using an eigenstructure assignment approach.
The proposed design method allows for several design objectives, i. e. we are considering non-overshooting and monotonic step-reference tracking of the output.
Additionally, we provide a structural condition for the solvability of the design problem for the switched systems.
This structured verification offers an additional dimension for evaluating the applicability of the developed controller concept in various contexts of switching linear systems.

In order to obtain stability we rectify the eigenstructure of the subsystems as proposed in \cite{HonGWTR2023}.
This imposes restrictions on the assignable eigenstructure which are strongly dependent on the constituent subsystems.
Employing the results of \cite{SchN2010} for non-overshooting step-reference tracking we portray the eigenstructure assignments that are feasible for the given set of subsystems.
For that matter, a partitioning is chosen that simultaneously assigns the modes of the subsystems to their mutual outputs.
We present conditions to check feasibility of such partitioning for the eigenstructure assignment problem.
Further, for a given choice of partitioning we obtain specific sets of eigenvalues as well as eigenvectors to be chosen from.
It turns out that in most cases the choice of eigenvalues is only mildly restricted, whereas the choice of eigenvectors is limited, depending on the chosen structure and eigenvalues to be assigned.
However in many cases these sets still provide enough freedom for design to shape the output dynamics as desired.

The manuscript is structured as follows.
The following section gives the class of switched systems as well as the design problem considered in this contribution.
Section~\ref{sec:EA_LTI} summarises preliminary results for the eigenstrucutre assignment for LTI systems and adapts the formulations for the scheme considered in the paper.
In Section \ref{sec:EA_switched} we address the eigenstructure assignment for switched linear systems.
Based on the approach in \cite{HonGWTR2023}, we derive conditions to solve the eigenstructure assignment problem ensuring stability for arbitrary switching and obtain sets with eigenvectors that are feasible for the eigenstructure rectification of the subsystems.
We present our main results on shaping the step-tracking response of the switched system in Section \ref{sec:trackingresponse}.
We propose a method to design non-overshooting step-tracking of the output for arbitrary switching as well as a monotonic response.
For both designs we present a constructive algorithm to achieve the desired performance.
In Section \ref{sec:example} we illustrate both design approaches by an example and discuss their properties.

\section{Problem Definition}

In  this section we introduce the system class considered and formulate the stabilisation  and reference tracking problems we address  in this paper. 

\subsection{Switched linear system}
Consider the MIMO LTI subsystems 
\begin{eqnarray}
	\Sigma_q: \
	\left\{ \begin{array}{lcr}
		\dot{x}(t) = A_q\,x(t)+B_q\,u(t),\;\; x(0)=x_0 \in \real^n, 
		\hfill\cr
		y(t)=C\,x(t)\hfill \end{array} \right.
	\label{eq:sys_q}
\end{eqnarray}
with $A_q\in\R^{n\times n}$ 
and $q\in\mathcal{I}=\{1,2\}$. 
The input matrices $B_q\in\R^{n\times m}$, $q\in\mathcal I$ have full column rank and the output matrix $C\in\R^{p\times n}$ has full row rank.
Note that the output matrix $C$ is common to both subsystems. 

These subsystems constitute the open-loop switched system 
\begin{align}\label{eq:sysOL}
	\Sigma_{\rm OL}: \
	\left\{ \begin{array}{lcr} \dot{x}(t)=A_{\sigma(t)}x(t)+B_{\sigma(t)}u(t)  \\
		y(t)=Cx(t), \hfill \end{array} \right. 
\end{align}
where the switching signal $\sigma :\real^{+}  \rightarrow \mathcal{I}$ indicates the active subsystem at time $t\geq 0$.
Here and in remainder of this paper the notation is to be understood as $A_{\sigma(t)}=A_q$ whenever $\sigma(t)=q$.
Note that the state vector~$x$ and output $y$ are common to all subsystems; only the system  dynamics and input matrices change at the switching instances.

We shall say  that $\sigma$ is an {\it  admissible switching signal} if  it is a 
piecewise constant function  with at most finitely many switches in any finite time interval. 
This guarantees the uniqueness  and  absolute continuity of the solution $x:\R\rightarrow \R^n$ of \eqref{eq:sysOL} for any initial condition $x_0\in\R^n$ and bounded input $u$.
Since the output matrix $C$ is common to both subsystems, also the output $y$ is absolutely continuous.

We firstly define the problem of  stabilising switched  linear systems  subject to  arbitrary switching. 
Next we define  our reference tracking problem for switched systems. Finally  we  describe the  problems  of  non-overshooting and monotonic  set-point tracking for switched systems,  and  introduce some necessary assumptions for these tracking  control problems to  be solvable.

\subsection{Stabilising state-feedback for switched linear systems}

Consider the homogeneous part of \eqref{eq:sysOL} given by
\begin{align} \label{eq:sys_sw_hom}
	\dot{x}(t)=A_{\sigma(t)}x(t), \quad x(0)=x_0\in\mathbb R^n.
\end{align}

\begin{defi}
	The  switched system \eqref{eq:sys_sw_hom} is globally uniformly asymptotically stable  under arbitrary switching (GUAS), if the origin is a uniformly asymptotically stable equilibrium point, for any admissible switching signal $\sigma$, and from all initial  states $x_0\in\mathbb R^n$.
\end{defi}

It is well known that \eqref{eq:sys_sw_hom} may have unbounded solutions for some switching signals $\sigma$, even when both system matrices $A_q$  are Hurwitz stable matrices \cite{LibMor99}.

For the control problem we shall assume that the switching of \eqref{eq:sysOL} occurs autonomously and cannot be influenced.
However we shall assume that current value $\sigma(t)$ of the switching signal is  known at any time instant $t\geq 0$ for the  purpose of controller design, and the control input $u$ can adjust to switches instantaneously.

Consider the controller synthesis problem  of   obtaining state 
feedback matrices $F_q \in \real^{m \times n}$,  $q\in\mathcal{I}$,   such  that the control law
\begin{align}
	u(t) =  F_{\sigma(t)}  x(t)
	\label{ulaw}
\end{align}
applied to  \eqref{eq:sysOL}, 
yields the  closed-loop switched system
\begin{align}
	\dot{x}(t)=(A_{\sigma(t)}+B_{\sigma(t)}F_{\sigma(t)})x(t),
	\label{eq:sys_closedloop_hom}
\end{align}
which is GUAS.

The system \eqref{eq:sysOL} is called \emph{stabilisable}, if there exist matrices $F_q \in \real^{m \times n}$,  $q\in\mathcal{I}$, such that \eqref{eq:sys_closedloop_hom} is GUAS for all admissible switching signals.

\subsection{Set-point tracking  for switched linear systems}

The  particular  problem addressed in this  paper is to design control laws that will achieve  stabilising set-point tracking under arbitrary switching,  while also  delivering a desirable transient response.

Given the switched system \eqref{eq:sysOL} and a constant reference signal  $r \in \real^p$, we consider the switched control law
\begin{align}
	u(t) = F_{\sigma(t)}x(t) + G_{\sigma(t)}
	\label{uqlaw}
\end{align}
with feedback matrices $F_q \in \real^{m \times n}$ and feedforward matrices $G_q\in \R^m$, $q\in\mathcal{I}$ to be chosen, that yields the closed-loop control system
\begin{align}  \Sigma_{\rm CL}\!:\! 
	\left\{ \begin{array}{lr}\dot{x}(t)=(A_{\sigma(t)}+B_{\sigma(t)}F_{\sigma(t)})x(t) + B_{\sigma(t)}G_{\sigma(t)}
		\hfill\cr
		y(t)=Cx(t).\hfill \end{array}\right. \label{eq:sysCL}
\end{align}

Our goal is to devise $F_q \in \real^{m \times n}$ and $G_q\in \R^m$, $q\in\mathcal{I}$, such that 
the state trajectories of  $ \Sigma_{\rm CL}$ converge to  an equilibrium point $x_\mathrm{ss}\in \real^n$ with  $y_\mathrm{ss}:=  Cx_\mathrm{ss} = r$.
Following \cite{SchN2010,NtoTSF2016} we shall additionally distinguish the transient characteristics of \emph{monotonic} and \emph{non-overshooting} output. Introducing the tracking error $e(t) := r - y(t) $ these notions are defined as follows.

\begin{defi}\label{def:nonovershoot}
The switched system $\Sigma_{\rm CL}$ achieves  \emph{non-overshooting tracking} of $r\in\R^p$ from the initial condition $x_0 \in \real^n$,
if the error converges, $e(t)\rightarrow 0$, and has no change of sign 
in any component, i.e. for each $k \in \{1,\ldots,p\}$, $\sgn(e_k(t))$ is constant for all $t >  0$. Here  $e_k$ denotes the  $k$-th component of $e$.
\end{defi}
\begin{defi}\label{def:monotonic}
The switched system $\Sigma_{\rm CL}$ achieves \emph{globally monotonic tracking} of $r\in\R^p$ 
if the error $e(t)\rightarrow 0$ converges monotonically in every output component, 
from \emph{any} initial condition $x_0\in\R^n$. 
\end{defi}

Thus our aim in this paper is given the switched system \eqref{eq:sysOL}, devise  matrices $F_q$ and  $G_q$ 
such that \eqref{eq:sysCL} is GUAS and also achieves
non-overshooting or monotonic set-point tracking under arbitrary switching signals $\sigma$ for any given $r\in\R^p$.
Formally we shall distinguish the following two design goals:

\begin{problem}{\bf Non-overshooting set-point tracking.}\label{prob:nonovershoot}\\
Given the switched system $\Sigma_{\rm OL}$ in \eqref{eq:sysOL} and the reference $r\in\R^p$.
Find  matrices $F_q$ and $G_q$ and a set $\mathcal X_0 \subset \real^n$ such that
$\Sigma_\mathrm{CL}$ in \eqref{eq:sysCL} is GUAS, 
and has non-overshooting tracking of $r$ from all $x_0\in\mathcal X_0$  under arbitrary admissible switching signals $\sigma$.
\end{problem}
\begin{problem}{\bf Global monotonic set-point tracking.}\label{prob:monotonic}\\
Given the switched system $\Sigma_{\rm OL}$ in \eqref{eq:sysOL} and the reference $r\in\R^p$.
Find  matrices $F_q$ and  $G_q$ such that 
$\Sigma_\mathrm{CL}$ in \eqref{eq:sysCL} is GUAS, 
and has globally monotonic tracking of $r$ under arbitrary admissible switching signals $\sigma$.
\end{problem}

\subsection{Common steady state and design of the feedforward control}

In this section we discuss some preliminary considerations regarding the set-point tracking control for switched systems and note some fundamental assumptions for our approach in this paper.

If the switched system is at an  equilibrium state of the active subsystem immediately  prior to switching,  the system  output may exhibit significant transients due to the switched dynamics. 
Such transients can be avoided if all subsystems share the same steady-state point~\cite{WulWirSho05}.
In this spirit we formulate the following assumptions.
\begin{ass} \label{Ass1} 
The dimensions  of  the  system matrices of $\Sigma_q$ satisfy  $n+p  < 2m$, each subsystem~$\Sigma_q$ is  right invertible \cite{Trentelman2002}, and the switched system $\Sigma_{\rm OL}$ is stabilisable. 
\end{ass}
Assumption \ref{Ass1} will be taken  as  a standing assumption  throughout the  paper, since without it we  have no prospect of   finding  feedback matrices $F_q$ that   render \eqref{eq:sys_closedloop_hom} GUAS.
With these assumptions we consider the following system of linear equations 
\begin{align}
\begin{bmatrix} A_1  &  B_1 & 0 \\
	A_2  & 0 &  B_2 \\
	C  & 0  &  0  \end{bmatrix}                                 
\begin{bmatrix}    x_{\rm ss} \\
	u_{1,\rm ss} \\
	u_{2,\rm ss} \end{bmatrix}
= 
\begin{bmatrix}    0 \\
	0 \\
	r \end{bmatrix},
\label{eq:x_ss_condition}
\end{align}
where the zero matrices are of appropriate dimensions.
For any given $r\in\R^p$ the solution of  \eqref{eq:x_ss_condition} yields the common steady-state $x_{\rm ss}$ for the subsystems $\Sigma_q$, $q \in\mathcal I= \{1,2\}$, producing the output $y=r$ with constant inputs $u_{1,\rm ss}$ and $u_{2,\rm ss}$, respectively.  Assumption \ref{Ass1} and the fact that $C$ has full row rank ensure such solutions  exist for any given $r\in\R^p$.

Assuming the given switched system \eqref{eq:sysOL} is stabilisable by matrices $F_q$, $q\in\mathcal I$,
we choose the feedforward matrices $G_q$, $q\in\mathcal I$, for the control law  \eqref{uqlaw}
\begin{align}\label{eq:Gk}
G_q = -F_q x_\mathrm{ss}  + u_{q,\mathrm{ss}}.
\end{align}

With the change of variable $\xi := x -x_{\rm ss}$ and the error term $e(t) = r - y(t) $ the control law~\eqref{ulaw} with feedforward \eqref{eq:Gk} yields the closed-loop homogeneous system
\begin{align}
\Sigma_\xi: 
\left\{ \begin{array}{lcr}
	\dot{\xi}(t)  = (A_{\sigma(t)}+B_{\sigma(t)}\,F_{\sigma(t)})\,\xi(t)\\ 
	e(t)  =  C\,\xi(t), \hfill \end{array} \right.
\label{eq:sys_xi}
\end{align}
equivalent to \eqref{eq:sysCL}.
If the state matrices $F_q$ achieve GUAS for \eqref{eq:sys_xi},  we conclude that $\xi$ converges to the  origin, the state vector  $x$ of $\Sigma_{\rm CL}$ converges to $x_{\rm ss}$, $e(t) \ra 0 $ and the  output  $y$ converges to $r$ as $t \ra \infty$.

The principal  task  will  be the design  of  the  feedback matrices $F_q$, $q\in\mathcal I$,  and the identification   of the  set $\mathcal X_0$ of initial conditions under which the control law \eqref{uqlaw} is able to  solve  Problems \ref{prob:nonovershoot}  and \ref{prob:monotonic}. 
The main  results  of this paper will  provide 
conditions for the solvability of  these  problems for a given switched system  $\Sigma_{\rm OL}$. When  the problems  are solvable,   we provide a constructive procedure to design controllers that solve the respective problems.
It turns out that the two properties may even be achieved in a mixed fashion such that Problem \ref{prob:nonovershoot} and \ref{prob:monotonic} apply to some chosen output components of the switched system only.

\section{Eigenstructure assignment for   LTI  MIMO  systems} \label{sec:EA_LTI}

In  this  section we  revisit  and  generalise  some earlier results on eigenstructure assignment by  state feedback for  LTI  systems. The  primary goal is  to obtain  a   closed-loop  eigenstructure  in which  only a  small number  of modes  (eigenvalues)  contribute to each  output.  In  this section we consider the following controllable, right-invertible LTI system
\begin{align} 
\begin{array}{lcr}
	\dot{x}(t) = A\,x(t)+B\,u(t),\qquad x(0)=x_0 \in \real^n, 
	\hfill\cr
	y(t)=C\,x(t)\hfill \end{array} 
\label{eq:sysLTI}
\end{align}
with $A\in\R^{n\times n}$, $B\in\R^{n\times m}$,  $C \in \real^{p \times n}$ and $p<m$.

Our analysis will require some tools from  the geometry of linear systems  \cite{Basile-M-92}. 
We define the {\it Rosenbrock matrix}  of \eqref{eq:sysLTI} as \begin{align}
\label{ros}
R(\lambda) \coloneqq \bmat{cc} \lambda\,I_n-A & B \\ C & 0 \emat, \quad \lambda \in \complex.
\end{align}
The invariant zeros of \eqref{eq:sysLTI} are the values of $\lambda \in \complex$ for which the rank of $R(\lambda)$ is strictly smaller than its normal rank.
We denote by  $\gV^*$  the {\it largest output-nulling subspace of \eqref{eq:sysLTI}}, i.e., the largest subspace $\gV$ of $\real^n$ for which a matrix $F\,{\in}\,\mathbb{R}^{m\,{\times}\,n}$ exists such that $(A+B\,F)\,\gV\subseteq \gV \subseteq \ker (C+D\,F)$. Any real matrix $F$ satisfying this inclusion is called a {\it friend \/} of $\gV$.
The symbol $\gR^*$ denotes the so-called {\em output-nulling reachability subspace} on $\gV^*$. It is the maximum  subspace reachable from the origin
with state trajectories completely belonging to $\gV^*$.

The aim of our eigenstructure assignment is  to render certain modes invisible from certain outputs.
For this matter we introduce the
matrices   $ C_{(k)}  \in \real^{p_{(k)} \times n} $ for the system \eqref{eq:sysLTI}, such that for each  $k \in \{0,1,\dots, p\}$,
\begin{align}
C_{(0)} & :=   C,  
\\
C_{(k)} & :=  C, \mbox{  with the $k$-th row removed.} \label{Ckdef2}
\end{align}
Accordingly, we have $p_{(0)} = p $ and $p_{(k)} = p-1$, $k\in\{1,\ldots,p\}$.
We will  use $\Sigma_{(k)}$ to denote the LTI system obtained from  the  triple $(A,B, C_{(k)})$,  with $\Sigma_{(0)}$ denoting the system $(A,B,C)$.   

\begin{remark} \label{Ck_remark}
The motivation  for introducing the matrices $C_{(k)}$ may  be described as   follows. The   outputs of the closed-loop system \eqref{eq:sys_xi} are determined by the eigenvectors of the closed loop-matrices $A_q + B_qF_q$. We aim to choose matrices $F_q$ such that these  eigenvectors are the same for   $q \in \mathcal I$,  and that,  for some $k$,  they  lie within the  kernel of $C_{(k)}$. 
This ensures that the $k$-th mode  contributes  only to the  $k$-th output (or else none  of the outputs, if it is in the  kernel of $C$).  By reducing the number of  eigenvalues/vector pairs that contribute to  each output  component,  we are able to constrain the shape of the output. If only a single  mode appears in each output,  the  response  is monotonic. If  two or at  most three  modes  contribute  to  each output, then the response will be non-overshooting from  a set of initial  conditions. We show in Section \ref{sec:trackingresponse} that this  set  can  be analytically determined,  and  to  some extent,  chosen  by selecting   suitable closed  loop eigenvalues. 
\end{remark}
\begin{defi}   \label{def:Rosenbrockk}  
For each  $\Sigma_{(k)}$, with $k \in \{0, 1,\dots, p\}$,   and  for  any  $\lambda \in \real$, we define the Rosenbrock system matrices 
\begin{align}
	\label{eq:Rosenbrockk}
	R_{(k)}(\lambda):=\begin{bmatrix}
		\lambda I-A &B\\C_{(k)} & 0
	\end{bmatrix}
\end{align}
of  dimension $(n+p_{(k)}) \times (n +m)$,  where  $p_{(0)} = p $ and $p_{(k)} = p-1$, $k\in\{1,\ldots,p\}$.  Their kernels  may  be decomposed as
\begin{equation}   \label{Roskernel}
	\im  \begin{bmatrix}
		N_{(k)}(\lambda)\\M_{(k)}(\lambda)
	\end{bmatrix}:=\ker(R_{(k)}(\lambda)),
\end{equation} 
where $N_{(k)}(\lambda) $  and  $M_{(k)}(\lambda)$ have dimensions $n \times m_{(k)}(\lambda)$ and  $m \times m_{(k)}(\lambda)$, respectively.
\end{defi}

\begin{remark}\label{rem:mk}
For any  $\lambda \in \complex $ that is  not an invariant  zero  of system  $\Sigma_{(k)}$,  the  rank of $R_{(k)}(\lambda)$ is the  {\it normal rank} of $\Sigma_{(k)}$,  denoted by $r_{(k)}$. As $B$ and $C_{(k)}$ are of full rank, we have $r_{(k)} =   n + p_{(k)}$,  and hence
$m_{(k)}(\lambda) = n +  m - r_{(k)} =  m -  p_{(k)}$. When $\lambda $ is  an invariant zero of  $\Sigma_{(k)}$,  we  have 
$\rk(R_{(k)}(\lambda) )  <  r_{(k)}$,  and hence $m_{(k)}(\lambda)  > m -  p_{(k)}$.

To  avoid unduly  cumbersome  notation,  we shall  assume for simplicity,  when referring to the dimensions of the  kernel of the  Rosenbrock  matrix  as in \eqref{Roskernel},  that $\lambda$ is  not  an  invariant zero. 
This will allow us to drop  the dependency of $m_{(k)}(\lambda) $ on $\lambda$,  and allow  us to  assume that 
$N_{(k)}(\lambda) $  and  $M_{(k)}(\lambda) $ have dimensions $n \times m_{(k)}$ and  $m \times m_{(k)}$, respectively,  where  $m_{(k)} = m - p_{(k)}$.
\end{remark}

\begin{remark} The utility of the Rosenbrock  matrix  for our Problems  \ref{prob:nonovershoot} and  \ref{prob:monotonic} can  be  understood a follows. 
Let $[v^\top\; w^\top]^\top \in \ker(R_{(k)}(\lambda)) \subset \real^{n+m}$,  with  $v\in\mathbb R^n$ and $w\in\mathbb R^m$.
Using the celebrated algorithm of  Moore \cite{Moo76}, e.g. Theorem \ref{thm:ea_LTI},  $v$ and $w$ can be used to obtain  a feedback matrix $F$ that ensures $v$ is an  eigenvector  of $A + BF$,  and also lies in the kernel of $C_{(k)}$,  as described in Remark \ref{Ck_remark}.
\end{remark}

The following notations and definitions will allow us  to summarise several  results  of \cite{SchN2010} in  a  succinct  manner.

\begin{defi}

\begin{enumerate}
	\item For any two positive  integers  $n_1<n_2$,  we  let $2^{[n_1, n_2]}$ denote the set of all subsets of the integers in the interval $[n_1, n_2]$.
	\item For  any  pair  of positive integers   $p, n$ with \mbox{$p<n$},  we say  that a $(p+1)$-tuple of integers    $(d_0, d_1, \dots, d_p)$ is  a \it{$(p,n)$-partitioning} if $0\leq d_0$, 
	$0 \leq d_k \leq 3$  for $ k\neq 0$, and   $\sum_{k =0}^{p} \ d_k = n$.
	\item  Let  $(d_0, d_1, \dots, d_p) $ be  a   $(p,n)$-partitioning.
	\begin{enumerate}
		\item Let  $\mathcal L_{k}=\{\lambda_{k,1},\ldots, \lambda_{k,d_k}\}\subset\R$, for each  $k\in\{0,1,\ldots,p\}$.  We say that the sets of scalars $\mathcal L_{k}$ are compatible  with 
		$(d_0, d_1, \dots, d_p) $ if $card(\mathcal L_k) =  d_k$   and they are  pairwise disjoint, i.e.  
		\begin{align*}
			\mathcal L_{k_1} \cap   \mathcal L_{k_2}  =\emptyset,
		\end{align*}
		for  any distinct $k_1, k_2 \in   \{0,1,\ldots,p\}$. We further assume that compatible  sets are indexed such that
		\begin{align*}
			\lambda_{k,1} < \lambda_{k,2}  < \lambda_{k,3}, \;\;k\in\{1,\ldots,p\}.
		\end{align*}
		\item Let  $\mathcal V_{k}=\{v_{k,1},\ldots, v_{k,d_k}\}\subset\R^n$, for each  $k\in\{0,1,\ldots,p\}$.  We say that the sets of vectors $\mathcal V_{k}$ are compatible  with 
		$(d_0, d_1, \dots, d_p) $ if $card(\mathcal V_k) =  d_k$. 
	\end{enumerate}
\end{enumerate} 
\end{defi}

\begin{remark}
A $(p,n)$-partitioning represents a desired allocation of  modes per output component.  The sets $\mathcal L_k$ represent  choices of desired closed-loop  eigenvalues,  and the vectors  in $\mathcal V_k$ represent the associated eigenvectors. 
Provided that $ \cup_{k=0}^p \mathcal V_k$  is  linearly  independent, the  methods of \cite{Moo76}  can be  used to obtain 
a feedback matrix $F$ such the  control law $u = Fx$  for  \eqref{eq:sysLTI}  ensures the   closed-loop  system has an eigenstructure in which the closed-loop eigenvalues in $\mathcal L_0$ will  not contribute to any of the  output components, and the modes in $\mathcal L_k$ will only  contribute to the  $k$-th output.  
The reason  for constraining $d_k \leq 3$ is that our design methods  seek  to  associate at most  three eigenvalue/eigenvector pairs  with each component. Then the error term $e_k$ for the $k$-th component  will  consist  of at  most  three exponential  terms. For a given initial condition, the error term can be obtained  in closed form,  and the Lemmas \ref{lemA1} and \ref{lemA2} in the Appendix can be used to determine if the error term  changes sign, and hence the response is overshooting.  These lemmas allow for at most three exponential terms in the error function. Corresponding lemmas when the sum of  four  or  more exponential  terms  change sign were given in \cite{SchN2010},  however these conditions are known to be highly  conservative, and hence will not be considered in our analysis.
\end{remark}

The parametrisation introduced above allows to combine several results on eigenstructure assignment from \cite{SchN2010} in the following theorem.
\begin{theorem} 
\label{thm:ea_LTI}
Given the LTI system \eqref{eq:sysLTI}, let the sets   $\mathcal{L}_k = \{\lambda_{k,1}, \ldots, \lambda_{k,d_k} \} \subset \real$, and 
$\mathcal V_k  = \{v_{k,1}, \ldots, v_{k,d_k} \} \subset \real^n$,  $k \in\{0,1,\ldots,p\}$, both  be compatible with  the  $(p,n)$-partitioning  $(d_0, d_1,\dots, d_p)$. Assume  $\mathcal V = \cup_{k=0}^p \mathcal V_k$ is  linearly  independent. Let $w_{k,i}\in\R^m$ be such that
\begin{align}
	\begin{bmatrix}  v_{k,i}  \\ w_{k,i}  \end{bmatrix}  \subset   \ker(R_{(k)}\big(\lambda_{k,i}) \big), \
	\label{MooreM2}
\end{align}
for each $\lambda_{k,i} \in \mathcal L_k$  and $v_{k,i} \in \mathcal V_k$.  
Let 
\begin{align}  F = -WV^{-1},  \label{MooreF}
\end{align} 
where $V \in \real^{n \times n}$, $W \in \real^{m \times n}$ are given  by
\begin{subequations}\label{MooreVW}
	\begin{align}  \label{eq:MooreV}
		V  &= \begin{bmatrix}
			v_{0,1} \ldots  v_{0,d_0} &v_{1,1} \ldots v_{p,d_p} 
		\end{bmatrix} ,\\
		W  &= \begin{bmatrix}
			w_{0,1} \ldots w_{0,d_0} &w_{1,1} \ldots  w_{p,d_p} 
		\end{bmatrix}.
	\end{align}  
\end{subequations} 
Then 
\textbf{ \begin{align}
		(A+B\,F)\,v_{k,i} & = \lambda_{k,i}\, v_{k,i},   \label{MooreF1} \\ 
		C_{(k)}\,v_{k,i}  & =  0.    \label{MooreF2}
\end{align}}
Defining 
\begin{align}\label{eq:def_alpha}
	[\alpha_{0,1}, \ldots, \alpha_{0,d_0} ,\;\alpha_{1,1}, \dots,  \alpha_{p,d_p}]^\top := V^{-1}x_0,
\end{align}
the output $y$ of \eqref{eq:sysLTI} from the input $u = Fx$  is given by
\begin{align}
	y(t)  =       \begin{bmatrix}          
		\alpha_{1,1} e^{\lambda_{1,1}t} + \dots  + \alpha_{1,d_1} e^{\lambda_{1,d_1} t} \\
		\alpha_{2,1} e^{\lambda_{2,1}t} + \dots  + \alpha_{2,d_2} e^{\lambda_{2,d_2} t} \\      
		\vdots\     \\
		\alpha_{p,1} e^{\lambda_{p,1}t} + \dots  + \alpha_{p,d_p} e^{\lambda_{p,d_p}t}  \label{eq:output_ea_LTI}
	\end{bmatrix}.
\end{align}
\end{theorem}%
\begin{proof}
For  each  $k\in\{0,1,\ldots,p\}$ and $i\in\{1,\ldots d_k\}$, if $\lambda_{k,i}, \ v_{k,i}  $  and $w_{k,i}$  satisfy \eqref{MooreM2},  and if $F$ is obtained from \eqref{MooreF}, then by Moore's  theorem \cite{Moo76}, 
$\lambda_{k,i}$ and  $v_{k,i}$ satisfy \eqref{MooreF1}-\eqref{MooreF2}. Thus 
\begin{align}\label{eq:Cvki}
	Cv_{k,i}  & = 
	\begin{cases}
		0 & \text{if } k=0, \\
		s_k & \text{if } k\in\{1,\ldots,p\},
	\end{cases}
\end{align}
for all $i\in\{1,\ldots,d_k\}$, where $0$ denotes the zero  vector and $s_k$ some multiple of the  $k$-th  canonical  basis vector of  $\real^p$. 
Hence the closed-loop output is 
\begin{align}
	y(t)  & =  C e^{(A+BF)t} x_0  
	=   \sum_{k=0}^{p} \sum_{i=1}^{d_k}\ C v_{k,i} \,\alpha_{k,i}\,e^{\lambda_{k,i}  t}, 
\end{align}
and with \eqref{eq:Cvki} we obtain the form \eqref{eq:output_ea_LTI}.

\end{proof}

\begin{remark}
The assigned closed-loop eigenstructure \eqref{MooreF1}-\eqref{MooreF2} ensures that  $d_0$ of  the closed-loop   eigenvectors are in the nullspace  of $C$. The remaining $n - d_0$ eigenvectors  are chosen 
so  that  exactly  $d_k$ of them lie  in the nullspace of $C_{(k)}$. Consequently, the  modes in $\mathcal L_0$ are rendered invisible at the outputs, and the modes in $\mathcal L_k$  are  visible  only in the  $k$-th output component,  for  each $k\in\{1,\ldots,p\}$.

Theorem \ref{thm:ea_LTI}  is a  combined restatement of  Theorems 3.1, 3.2  and  3.3 in \cite{SchN2010}. The  presentation is also  somewhat more flexible, since it  allows  for  one,  two or three modes to  be  allocated to   each output component. The three theorems in \cite{SchN2010} assumed each output would  have exactly  one ($d_k \equiv 1$), or else  exactly  two ($d_k \equiv 2$),  or exactly three  ($d_k \equiv 3$),  modes per  output. 
\end{remark}

Theorem \ref{thm:ea_LTI} requires the existence  of a $(p,n)$-partitioning, compatible sets $\mathcal L_k$ and  linearly independent $\mathcal V_k$.
Some necessary conditions for this can   readily be obtained from  the  perspective  of linear systems geometry. The   $d_0$ vectors in $\ker(C)$ must  lie within $\mathcal R^*$, the largest  output-nulling reachable  subspace of the system  $(A,B,C)$   \cite{BasM.1992}.  The feedback matrix  $F$ is  a friend  of this subspace.
If we let   $\mathcal R_{(k)}^*$ denote the  largest   output-nulling reachable  subspace of system $(A,B,C_{(k)})$,  then \eqref{MooreM2}  for  \mbox{$k\in\{0,1,\ldots,p\}$},  requires that for  any closed loop eigenvalue-eigenvector pair $(\lambda_{k,i}, v_{k,i})$   with $\lambda_{k,i} \in \mathcal L_k$,  we  must have  $v_{k,i}\in \mathcal R_{(k)}^*$.
Thus a  necessary condition for the existence of a  friend that assigns the desired eigenstructure \eqref{MooreF1}-\eqref{MooreF2}  is that   $\dim(\mathcal R^*) \geq   d_0$, 
and $\dim(\mathcal R_{(k)}^*) \geq d_0 + d_k$,  since $\mathcal R^*  
\subset \mathcal R_{(k)}^*$,  and the vectors $d_k$ selected from  $ \mathcal R_{(k)}^* $  must be independent from the $d_0$ vectors selected  from $\mathcal R^*$.
We define
\begin{defi}  A $(p,n)$-partitioning  $(d_0, d_1,\dots, d_p)$   is feasible for $(A,B,C)$ in \eqref{eq:sysLTI} if 
\begin{align}
	d_0 & \leq  \dim(\mathcal R^*)  \label{dimRstar} 
	\\
	d_{k} &  \leq \dim(\mathcal R_{(k)}^*) -  d_0.\label{dimRstar_k}
\end{align}   
\end{defi}

Note that Conditions \eqref{dimRstar} and \eqref{dimRstar_k} constitute necessary but not sufficient conditions for solving the problems addressed in this paper.
In this sense such partitioning is a feasible candidate for pursuing the considered design goals.
Conditions \eqref{dimRstar} and \eqref{dimRstar_k} are readily checked using the MATLAB GA toolbox \cite{BasM.1992}.

Assuming  at least  one feasible partitioning exists,  the task is to identify   
conditions  ensuring the existence of suitable  sets $\mathcal L_k$ and  $\mathcal V_k$. We formally state  this as: 

\begin{problem} {\bf Eigenstructure assignment for LTI  systems.} \label{Prob:estruct1}
Let the $(p,n)$-partitioning $(d_0, d_1,\dots, d_p)$ be feasible for \eqref{eq:sysLTI}, and  let the sets $\mathcal{L}_k $ and  $\mathcal V_k$, $k\in\{0,1,\ldots,p\}$,  be  compatible with $(d_0, d_1,\dots, d_p)$. Find (if possible) a feedback matrix $F$ that  yields  a closed-loop output given  by 
\eqref{eq:output_ea_LTI}.
\end{problem}

To  obtain  sufficient conditions for solving this  problem,  we introduce, for a  given $\lambda \in \real$ and  for each   $k \in\{0,1,\ldots,p\}$, the subspaces
\begin{align}
\hspace{-0.5em}\mathcal{R}_{(k)}^*(\lambda) \!\! := \!\! \left\{\! v \in \real^n: \exists w \in \real^m \left| 
\begin{bmatrix}
	v  \\ w
\end{bmatrix}
\!\in\! \ker\big(R_{(k)}(\lambda) \big) \!\right\} \right.   \label{Rstarlambda}\!,
\end{align}
and for a given set $\mathcal L_k\subset \R$ that is compatible with a  given feasible $(p,n)$-partitioning,
\begin{align}
\mathcal{ R}_{(k)}^* ( \mathcal  L_k  ) :=  \sum_{ \lambda \in \mathcal   L_k}  \mathcal{ R}_{(k)}^*(\lambda).  \label{RstarL}
\end{align}
Thus $\mathcal{R}_{(k)}^*(\lambda) $ represents the closed-loop eigenvectors that that can be assigned to satisfy \eqref{MooreF1}-\eqref{MooreF2},  and 
$\mathcal{ R}_{(k)}^* ( \mathcal  L_k  ) $  is the subspace of  $\real^n$ that  is generated by  these vectors,   for all $\lambda  \in \mathcal L_k$. Our  solution  to Problem \ref{Prob:estruct1} is given by
\begin{theorem} \label{Prob1_solved}
Let the $(p,n)$-partitioning $(d_0, d_1,\dots, d_p)$ be feasible for \eqref{eq:sysLTI}, and  let the sets $\mathcal{L}_k = \{\lambda_{k,1}, \ldots, \lambda_{k,d_k} \} \subset \real$, $k\in\{0,1,\ldots,p\}$ be  compatible with $(d_0, d_1,\dots, d_p)$. Then a  feedback   matrix $F$ solving Problem \ref{Prob:estruct1} exists if,  for  every $S \in 2^{[1, p]}$,
\begin{align}
	\dim\bigg(\mathcal{ R}_{(0)}^*( \mathcal  L_0 )
	+ \sum_{k \in S} \mathcal{ R}_{(k)}^*( \mathcal  L_k )  \bigg)
	= d_0 + \sum_{k \in S} \ d_{k}.    \label{Rkcond}
\end{align}
\end{theorem}
\begin{proof}
We apply Proposition  \ref{prop:rado} with subspaces $\mathcal P_k= \mathcal{ R}_{(k)}^* ( \mathcal  L_k  ) $. 
We obtain $n$ linearly independent vectors
$v_{k,i}$,  $k\in\{0,1,\ldots,p\}$ and $i\in\{1,\ldots,d_k\}$,
such that each $v_{k,i}  \in  \mathcal{ R}_{(k)}^* ( \mathcal  L_k  )$.
Solving \eqref{MooreM2} for each $v_{k,i}$, we obtain $n$ vectors $w_{k,i}$.
Since 
$v_{k,i}$ are linearly independent, from Theorem \ref{thm:ea_LTI}, we obtain $F$  in \eqref{MooreF}  such that \eqref{MooreF1}-\eqref{eq:output_ea_LTI} are satisfied.
\end{proof}

\begin{remark}    
\cite{NtoTSF2016}  gives necessary and sufficient results for the existence of a  feedback matrix $F$ to  solve the eigenstructure  Problem \ref{Prob:estruct1}  for the case  of monotonic outputs, in  which a  single  mode  contributes to  each output component. This  corresponds to  the  $(p,n)$-partitioning  with 
$d_0 = n-p$ and  $d_k =  1$, for all $k\in\{1,\ldots, p\}$. Theorem  \ref{Prob1_solved} generalises this result to accommodate an arbitrary  feasible $(p,n)$-partitioning for \eqref{eq:sysLTI}.
\end{remark}

\section{Eigenstructure assignment for switched linear systems} \label{sec:EA_switched}

In  this section  we revisit  the  eigenstructure results of \cite{HonGWTR2023}  on the global stability of LTI systems under arbitrary switching sequences,   and extend them  to include  systems  with outputs.
\begin{defi}\label{def:rectifiable}
For the  systems $\Sigma_q$ in \eqref{eq:sys_q},   if there exists  a set of linearly independent vectors $\mathcal V \! =\!  \{v_i\!: i \in\{1,\ldots n\}\} \subset \real^n$, and sets
$\mathcal L_q = \{\lambda_{q,i}: i \in\{1,\ldots n\}\} \subset \real$, with $q \in \mathcal{I}$,  
and matrices  $F_q\in \R^{m\times n}$ such that 
\begin{align}
	\label{eq:def_FBrect}
	(A_q+B_qF_q)v_i=&\lambda_{q,i} v_i 
\end{align} 
for each $i \in  \{1, \ldots, n\}$,  then  we say that 
the systems $\Sigma_{q}$  are  \it{simultaneously eigenvector rectifiable}.
\end{defi}

Note that for  simultaneous eigenvector  rectification,  we require the feedback matrices $F_q$ to assign  identical eigenvectors in ${\mathcal V}$ for both closed-loop subsystems, however the corresponding eigenvalues in   $\mathcal L_q  $ may be   different. The following result  from~\cite{HonGWTR2023} establishes that simultaneous eigenvector rectification with stable eigenvalues implies  GUAS.

\begin{theorem} \label{thm:GASAS}
Assume that  state feedback matrices  $F_q\in \R^{m\times n}$
achieve  simultaneous eigenvector  rectification  for $\Sigma_q$,  for some $\mathcal V \subset \real^n$  
and some stable sets  of  eigenvalues
$\mathcal L_q   \subset \real^-$,   $q \in \mathcal{I}$. Then $\Sigma_{\rm CL}$ is stable  under arbitrary switching sequences.
\end{theorem}

\begin{remark}
An important contribution of \cite{HonGWTR2023}  is to obtain  conditions on  $\Sigma_{\rm OL}$ ensuring that simultaneous eigenvector rectification  can  be achieved. 
The  system dimension  requirement $ n < 2m $ is shown to be  necessary, which is accommodated in our Assumption \ref{Ass1}.
\end{remark}

Our main task in  this section is to  extend the eigenvector rectification methods
to accommodate systems with outputs, while also  constraining  the  outputs to contain only   a few  closed-loop modes, as in Problem \ref{Prob:estruct1}.
We begin  by extending the  notation of Section \ref{sec:EA_LTI}   to the   systems $\Sigma_q$   in \eqref{eq:sys_q}. 

\begin{notat}        \label{notat1}
For each   $q \in \mathcal{I}$  and  $k \in \{0,1,\ldots, p\}$,  we  define $\Sigma_{q,(k)}  = (A_q, B_q, C_{(k)})$   with  $C_{(k)}$ defined as in \eqref{Ckdef2}.  For convenience  we identify 
$\Sigma_{q,(0)}  = \Sigma_{q}$ in this notation.
We  let   $ \mathcal R_{q,(k)}^*$ denote the largest   output-nulling reachable  subspace of  $\Sigma_{q,(k)}$ and where convenient,  we  identify   $\mathcal R_q^* =    \mathcal R^*_{q,(0)}$.  For any $\lambda \in \real$, we define the Rosenbrock system matrices  $R_{q,(k)}(\lambda)  $ 
according to \eqref{eq:Rosenbrockk} of  Definition \ref{def:Rosenbrockk}. Their kernels  may  be decomposed as $   N_{q,(k)}(\lambda) $  and 
$M_{q,(k)}(\lambda) $  in \eqref{Roskernel}.
Let $\mathcal L_{q,k} = \{\lambda_{q,k,1},\ldots,\lambda_{q,k,d_k}\}\subset\R$, $k\in\{0,1,\ldots,p\}$ be sets compatible with $(d_0,d_1,\ldots,d_k)$ and indexed such that
\begin{align}\label{eq:indexed_lambda_q}
\lambda_{q,k,1}\leq \lambda_{q,k,2}\leq \lambda_{q,k,d_k},\;\;k\in\{1,\ldots,p\}.
\end{align}
The set of closed-loop eigenvalues in Theorem \ref{thm:GASAS} is then obtained by $\mathcal L_q=\mathcal L_{q,0}\cup L_{q,1} \cup \cdots \cup L_{q,p}$.
\end{notat}%

\begin{remark}
To simplify notation (see also Remark~\ref{rem:mk}) we shall assume that $\lambda$ is not an invariant zero of $\Sigma_{q,(k)}$, when referring to the dimensions of $N_{q,(k)}(\lambda)$. 
Then $N_{q,(k)}(\lambda)\in\mathbb R^{n\times m_{(k)}}$ for $q\in\{1,2\}$.
Note however, that this assumption is made to simplify notation only, and should not restrict the choice of $\lambda$ in the design.
In fact if $\lambda$ is chosen to be an invariant zero of $\Sigma_{q,(k)}$, the dimension of $N_{q,(k)}(\lambda)$ is increased and thus may provide more choice for the joint eigenvector assignment that follows in \eqref{eq:intersection}.
\end{remark}

In  terms  of this   notation,  we  can  now  formalise the eigenstructure assignment problem to  be solved in order to achieve  non-overshooting  set-point tracking under arbitrary switching for  $\Sigma_{\rm CL}$ as follows:

\begin{problem} {\bf Eigenstructure assignment for switched linear systems with outputs.} \label{Prob:estruct2}
Given a  $(p,n)$-partitioning $(d_0, d_1,\dots, d_p)$ with  compatible  sets $\mathcal{L}_{q,k} = \{\lambda_{q,k,1}, \ldots, \lambda_{q,k,d_k} \} \subset \real$, $q\in\mathcal I$,  and  $\mathcal V_k \subset \real^n $, 
$k\in\{0,1,\ldots,p\}$, find   (if possible)   feedback matrices  $F_q$  for the feedback laws   $u=F_qx$ to achieve simultaneous eigenvector rectification \eqref{eq:def_FBrect}  and also yield  closed-loop outputs, for  both $\Sigma_q$, in  the form \eqref{eq:output_ea_LTI}.
\end{problem}

Note that the desired  eigenvalue sets $\mathcal L_{q,k}$ may be different for the two subsystems, however the $(p,n)$-partitioning  must be the same, in order to obtain outputs with the same  number of modes in each  component. 

All assignable eigenvectors for each $k \in \{0,1, \ldots,  p\}$ can be found in the following intersection
\begin{align}\label{eq:intersection}
v_{k,i}\in \im N_{1,(k)}(\lambda_{1,k,i})\cap\im N_{2,(k)}(\lambda_{2,k,i})
\end{align} for $i \in \{1,\ldots,d_k\}$. 

In order to describe the intersection \eqref{eq:intersection} as a span of a rational matrix-valued function we consider the 
scalars $\lambda_1, \lambda_2 \in \real$.  Let  $T_{(k)}(\lambda_1,\lambda_2)$ be a  transformation   matrix that converts $[ N_{1,(k)}(\lambda_1)\ N_{2,(k)}(\lambda_2) ] $  into reduced row-echelon  form:
\begin{align}\label{eq:rref}
T_{(k)}(\lambda_1,\lambda_2)\begin{bmatrix}N_{1,(k)}(\lambda_1)&N_{2,(k)}(\lambda_2)
\end{bmatrix}   =   \begin{bmatrix}
E_{(k),11}(\lambda_1,\lambda_2)&E_{(k),12}(\lambda_1,\lambda_2)\\0&E_{(k),22}(\lambda_1,\lambda_2)\\0&0
\end{bmatrix},
\end{align}
where $E_{(k),11}, E_{(k),12}\in\R[s]^{m_{(k)}\times m_{(k)}}$, and the zero-row has the dimension $p_{(k)}\times 2m_{(k)}$, since $C_{(k)}N_{q,(k)}(\lambda_q)=0$, $q\in\mathcal I$, i.e. the columns of all $N_{q,(k)}$ are orthogonal on $C_{(k)}$.
Define the set 
\begin{align*}
G_{(k)} := \big\{(\lambda_1,\lambda_2): T_{(k)}(\lambda_1,\lambda_2) \mbox{ is invertible} \big\} \subset \R\times\R.
\end{align*}
Adapting the result in \cite{HonGWTR2023} we obtain the following.
\begin{prop}\label{prop:Qk}
For any $(\lambda_1,\lambda_2) \in G_{(k)}$, let   $Q_{(k)}(\lambda_1,\lambda_2)$  be the rational matrix-valued function given  by
\begin{align}  \label{eq:def_Qk}
Q_{(k)}(\lambda_1,\lambda_2)= T_{(k)}^{-1}(\lambda_1,\lambda_2)\begin{bmatrix}
E_{(k),12}(\lambda_1,\lambda_2)\\0\\0
\end{bmatrix},
\end{align}
where $E_{(k),12}(\lambda_1,\lambda_2)$ is obtained by the reduced row-echelon form of $[ N_{1,(k)}(\lambda_1)\ N_{2,(k)}(\lambda_2) ]$ 
in \eqref{eq:rref}. Then 
\begin{align}\label{eq:assign_intersection_2_Q}
\im N_{1,(k)}(\lambda_1)\cap \im N_{2,(k)}(\lambda_2)=\im Q_{(k)}(\lambda_1,\lambda_2)\ . 
\end{align} 
\end{prop}

\begin{remark}
\label{rem:DifRep}
The sets  $G_{(k)}$ are  dense in $\mathbb R^2$ \cite{HonGWTR2023}, and
denote the subset of assignable eigenvalues for which  $\im Q_{(k)}(\lambda_1,\lambda_2)$ is a valid representation of the intersection  $\im N_{1,(k)}(\lambda_1)\cap \im N_{2,(k)}(\lambda_2)$.
\end{remark}
\begin{remark}\label{rem:defect}
In case $T_{(k)}(\lambda_1,\lambda_2)$ is singular at points of interest $(\lambda_1^\star,\lambda_2^\star)$, 
further representations of the intersection can be obtained.
At such points $(\lambda_1^\star,\lambda_2^\star)$ the reduced row-echelon form takes a different shape and is obtained by a different transformation $T_{(k)}^\star(\lambda_1^\star,\lambda_2^\star)$.
Calculating $Q^\star_{(k)}(\lambda_1^\star,\lambda_2^\star)$ may increase the number of assignable linearly independent eigenvectors, see \cite{HonGWTR2023} for an example.
\end{remark}

The set $G_{(k)}$ provides pairs  of suitable  eigenvalues for  solving Problem \ref{Prob:estruct2}; the next  task is to obtain a suitable $(p,n)$-partitioning and  compatible sets  of  eigenvalues.  
We define,  for each   $k\in\{0,1, \dots, p\}$, 
\begin{align} \label{eq:def_d(k)}
d_{(k)}  \coloneqq  \dim \left(\sum_{(\lambda_1,\lambda_2)\in G_{(k)}}\im Q_{(k)}(\lambda_1,\lambda_2)\right).
\end{align}

Thus $d_{(k)}$ represents the number of linearly independent  eigenvectors that can be assigned in $\im Q_{(k)}(\lambda_1,\lambda_2)$ for distinct $\lambda_1=\lambda_{1,k,i}\in\mathcal L_{1,k}$, $\lambda_2=\lambda_{2,k,i}\in\mathcal L_{2,k}$,
and $\mathcal L_{1,k}\times\mathcal L_{2,k}\subset G_{(k)}$.
To obtain a finite spanning set for the set in \eqref{eq:def_d(k)}, we obtain a polynomial representation of the image of $Q_{(k)}(\lambda_1,\lambda_2)$ by multiplying each column with 
the largest common denominator polynomial $q_{j} (\lambda_1,\lambda_2),\ j\in\{1,\ldots, m_{(k)}\}$ 
{\small
\begin{align}
\label{def_P_from_Q}
P_{(k)}(\lambda_1,\lambda_2)\!=\!Q_{(k)}(\lambda_1,\lambda_2)
\!\!\begin{bmatrix}
	q_1(\lambda_1,\lambda_2)&&0\\[-0.5ex]
	&\!\!\!\!\!\!\!\!\ddots\!\!\!\!\!\!\!\!&\\[-0.5ex]
	0& &q_{m_{(k)}}\!(\lambda_1,\lambda_2)\!
\end{bmatrix}\!\!.
\end{align}}%
$P_{(k)}(\lambda_1,\lambda_2)$ can be written as  a matrix polynomial with coefficient matrices $D_{(k),hl} \in \real^{n \times m_{(k)}}$
\begin{align}
\label{def:P}
P_{(k)}(\lambda_1,\lambda_2)=\sum_{h=0}^n\sum_{l=0}^n D_{(k),hl}\lambda_1 ^h\lambda_2^l\,.
\end{align} 
The following lemma permits the  computation of $d_{(k)}$.

\begin{lemma}
For each $k\in\{0,1, \dots, p\}$, let $D_{(k),hl}$ be the coefficient matrices in \eqref{def:P}.
Then
\begin{align}
\label{eq:dkviarank}
d_{(k)} \!=\! \rk\! \begin{bmatrix}   D_{(k),00} &\ldots & D_{(k),hl}&\ldots & D_{(k),nn}
\end{bmatrix}.
\end{align}

\end{lemma} 
\begin{proof}
Let $\tilde D_k=[\tilde D_{(k)00}\ \ldots\ \tilde D_{(k)nn}]$ consist of all linearly independent rows of
\begin{align*}
\begin{bmatrix}   D_{(k),00} &\ldots & D_{(k),hl}&\ldots & D_{(k),nn}
\end{bmatrix},
\end{align*}
and  denote $\tilde d_k = \rk \tilde D_k$. Define 
\begin{align}
\label{def:tildeP}
\tilde P_k(\lambda_1,\lambda_2)=\sum_{h=0}^n\sum_{l=0}^n \tilde D_{(k)hl}\lambda_1 ^h\lambda_2^l\,.
\end{align} 
Assume there exists  $\alpha\in\R^{\tilde d}\setminus\{0\}$ satisfying $\alpha^\top \tilde P_k(\lambda_1,\lambda_2)=0$ for all $(\lambda_1,\lambda_2)\in\R\times\R$. 
Then with \eqref{def:tildeP} we get
\begin{align}
\label{eq:lincomzero}
\alpha^\top\!\tilde P_k(\lambda_1,\lambda_2)\!=\!\sum_{h=0}^n\sum_{l=0}^n\alpha^\top\!\tilde D_{(k)hl}\lambda_1^h\lambda_2^l=0\in\R^{1\times z}.
\end{align}
Further, in the ring $\R[\lambda_1,\lambda_2]$ over $\R$, the set of polynomials $\{\lambda_1 ^h\lambda_2^l\}_{h,l=0}^n$ is linearly independent. Hence \eqref{eq:lincomzero} implies 
\begin{align*}
\alpha^\top\tilde D_{(k)hl}=0\quad \text{for all $h,l \in \{1,\ldots,n\}$},
\end{align*}
but this contradicts \eqref{eq:dkviarank}. 
Hence, there exists no $\alpha\in\R^{\tilde d}\setminus\{0\}$ satisfying $\alpha^\top\tilde P_k(\lambda_1,\lambda_2)=0$ for all $(\lambda_1,\lambda_2)\in\R\times\R$.
This implies that 
\begin{align*}
\dim \left(\sum_{(\lambda_1,\lambda_2)\in G_{(k)}} \im \tilde P_k(\lambda_1,\lambda_2)\right)=\tilde d_k 
\end{align*}
for any dense set $G_{(k)}\subset \R\times\R$.
With \eqref{def_P_from_Q} we have $\tilde d_k =d_{(k)}$.
\end{proof}

\begin{remark}
As the considered intersection is valid for all pairs of eigenvalues $(\lambda_1,\lambda_2)\in G_{(k)}$ it is possible to select any eigenvalues within this set in order to obtain $d_{(k)}$ linearly independent eigenvectors, as it is known that different eigenvalues always lead to different eigenvectors.
\end{remark}

\begin{defi}\label{def:feas_swsys}
A $(p,n)$-partitioning $(d_0, d_1,\dots, d_p)$   is feasible for the switched system $\Sigma_\mathrm{OL}$ in~\eqref{eq:sysOL} if 
\vspace{-0.5ex}
\begin{align}\label{eq:feas_swsys}  
d_k  \leq d_{(k)},\quad k\in\{0,1,\ldots, p\}.
\end{align}
\end{defi}

\begin{remark}
Note that Inequality \eqref{eq:feas_swsys} requires that the partitioning is feasible for both subsystems, but also requires that the intersections of each partition \eqref{eq:intersection} provide at least the requested number of linearly independent eigenvectors for each $k\in\{0,1,\ldots, p\}$.  
\end{remark}

The following  lemma is straightforward:
\begin{lemma}
A feasible  $(p,n)$-partitioning $(d_0, d_1,\dots, d_p)$   exists for  system $\Sigma_\mathrm{OL}$ if 
\begin{align}
\sum_{k=1}^p \  \min(3,d_{(k)}) \geq    n -  d_{(0)}.
\end{align}  
\end{lemma}

\begin{notat}Given a  feasible $(p,n)$-partitioning $(d_0, d_1,\dots, d_p)$  for $\Sigma_\mathrm{OL}$ with  compatible  sets $\mathcal{L}_{q,k} = \{\lambda_{q,k,1}, \ldots, \lambda_{q,k,d_k} \} \subset \real$, $q\in\mathcal I$, $k\in\{0,1,\ldots,p\}$, we  define
\begin{align}\label{eq:sum_of_Pks}
\mathcal{P}_{(k)} ( \mathcal L_{1,k}\!\times\! \mathcal L_{2,k}  )  :=\!\!  \sum_{ (\lambda_1,\lambda_2) \in \mathcal L_{1,k}\!\times\! \mathcal L_{2,k}} \!\!  \im P_{(k)}(\lambda_1,\lambda_2)  .
\end{align}
\end{notat}
Our solution to Problem  \ref{Prob:estruct2} is the main result of this section. It provides a sufficient condition for the simultaneous eigenvector rectification together with the simultaneous eigenvector assignment to the outputs of a switched linear system.
\begin{theorem}
\label{thm:Switched_Prob2_solved}
Let the $(p,n)$-partitioning  $(d_0, d_1,\dots, d_p)$ be feasible for $\Sigma_\mathrm{OL}$ with compatible  sets $\mathcal{L}_{q,k} = \{\lambda_{q,k,1}, \ldots, \lambda_{q,k,d_k} \} \subset \real$, $q\in\mathcal I$, $k\in\{0,1,\ldots,p\}$,
such  that $\mathcal L_{1,k}\times  \mathcal L_{2,k}\subset G_{(k)}$.
Then   feedback   matrices $F_q$ solving Problem \ref{Prob:estruct2} exist if
\begin{align}
	\hspace{-.5em}\dim\! \bigg(\!\mathcal P_{(0)}(\mathcal{L}_{1,0}\!\!\times\!\!\mathcal{L}_{2,0})\!
	+\! \sum_{k \in S}\! \mathcal P_{(k)}( \mathcal{L}_{1,k}\!\!\times\!\!\mathcal{L}_{2,k} ) \! \bigg)\!
	= \!d_0 \!+\!\! \sum_{k \in S} d_{k}.     
	\label{eq:test_overallfeasibility}
\end{align}
for every $S \in 2^{[1, p]}$.
\end{theorem}

\begin{proof}
Identify the subspaces $\mathcal P_k$ in Proposition \ref{prop:rado} with $\mathcal P_{(k)}( \mathcal{L}_{1,k}\!\!\times\!\!\mathcal{L}_{2,k} )$ of the theorem.
We obtain $n$ linearly independent vectors
$v_{k,i}$,  $k\in\{0,1,\ldots,p\}$ and $i\in\{1,\ldots,d_k\}$,
such that each $v_{k,i}  \in  \mathcal P_{(k)}( \mathcal{L}_{1,k}\!\!\times\!\!\mathcal{L}_{2,k} )$.
Solving \eqref{MooreM2} for each $v_{k,i}$ we obtain $n$ vectors $w_{q,k,i}$  for each subsystem $q\in\mathcal I$.
Since 
$v_{k,i}$ are linearly independent, we obtain $F_q=W_q V^{-1}$ with $V$ in \eqref{eq:MooreV} and $W_q= 
\begin{bmatrix}
w_{q,0,1} \ldots w_{q,0,d_0} &w_{q,1,1} \ldots  w_{q,p,d_p} 
\end{bmatrix}$.
Then \eqref{eq:def_FBrect} is satisfied for each $q\in\mathcal I$, and with Theorem \ref{thm:ea_LTI} the outputs of each subsystem $\Sigma_q$ with feedback $u=F_qx$ take the form \eqref{eq:output_ea_LTI}.
\end{proof}

\section{Main Results: Shaping of the tracking response}\label{sec:trackingresponse}

In this section we present our main results on shaping  the transient response  of the switched system for constant reference tracking. The main  theorems  solve Problems \ref{prob:nonovershoot}  and  \ref{prob:monotonic}. 
Moreover we give a constructive procedure for the synthesis of the switched control law \eqref{uqlaw} that achieves these desired  transient responses. 

The first result yields non-overshooting tracking of a  step response from a  set 
of initial states, under arbitrary switching. 
The second result achieves globally  monotonic tracking of constant references 
under arbitrary switching,  and requires stronger conditions on the properties of the subsystems.
For both results we obtain feedback matrices that achieve simultaneous eigenvector rectification, while also  assigning the closed-loop eigenvectors into suitable output-nulling subspaces.
Asymptotic stability of the closed-loop switched system is guaranteed by assigning eigenvalues in $\R^-$ for each mode.

For the  nonovershooting  design, we begin  by  identifying  a  set of  initial states from which the non-overshooting response will  be achieved. This requires the following  notation.
\begin{notat}
Let $(d_0, d_1,\dots, d_p)$  be a  feasible  $(p,n)$-partitioning  for $\Sigma_\mathrm{OL}$,  and let $\mathcal V_k = \{v_{k,1}, \dots, v_{k,d_k}\}$    be compatible sets of  vectors. We define
\begin{align}\label{eq:def_setH0}
\mathcal H_0 := \Span\Big\{v_{0,1}, \dots, v_{0, d_0}\Big\}
\end{align} 
and  for $k  \in \{1, \dots, p\}$, we define the sets 
\begin{align}\label{eq:def_setHk}
\mathcal H_k := \Big\{\xi\in\R^n\!:\xi=\sum_{i=1}^{d_k}\alpha_{k,i}\, v_{k,i} \Big\},
\end{align}
with $\alpha_{k,i} \in \real$. If $d_k=2$, the  $\alpha_{k,i} $ satisfy 
\begin{align}
(\alpha_{k,1} + \alpha_{k,2})\alpha_{k,2} > 0 ;   
\end{align}
if $d_k=3$,  the $\alpha_{k,i} $ are such  that  none of the following hold:
\begin{itemize}
\item[I.] $\alpha_{k,1} \, \alpha_{k,2}\!>\!0$, $\alpha_{k,1} \,\alpha_{k,3}\!<\!0$, and $|\alpha_{k,1}\! +\! \alpha_{k,2}|\! >\!
|\alpha_{k,3}|$;\\[-1.5ex]
\item[II.] $\alpha_{k,2} \, \alpha_{k,3}\!>\!0$,
$\alpha_{k,1} \, \alpha_{k,2}\!<\!0$, and $|\alpha_{k,1}|\!>\!|\alpha_{k,2}\!+\!\alpha_{k,3}|$;\\[-1.5ex]
\item[III.] $(\alpha_{k,2}+\alpha_{k,3})\alpha_{k,3}<0$. 
\end{itemize}%
We  also  let $\pi_k(\xi)$, $k\in\{0,1,\ldots,p\}$ denote the orthogonal projection of $\xi\in\R^n$ onto $\Span\{ v_{k,1},\ldots,v_{k,d_k}\}$.
\end{notat}

The following theorem provides the synthesis of the switched controller to obtain non-overshooting  set-point tracking.

\begin{theorem}[Non-overshooting outputs]
\label{thm:Final_nous}
Let the  switched system $\Sigma_{\rm OL}$  in \eqref{eq:sysOL} satisfy Assumption \ref{Ass1} and let $r \in \real^p$ be the desired set-point. Let the $(p,n)$-partitioning $(d_0, d_1,\dots, d_p)$ be  feasible for $\Sigma_{\mathrm{OL}}$,  and let  $\mathcal L_{q,k} \subset \R^-$ be  compatible  sets of desired  closed-loop  eigenvalues such  that  ${\mathcal L}_{1,k}  \times {\mathcal L}_{2,k} \subset G_{(k)}$,  and satisfy \eqref{eq:test_overallfeasibility}.  Let $\mathcal V_k $ be  compatible sets  of vectors  with 
$v_{k,i}\in\im P_{(k)}(\lambda_{1,k,i},\lambda_{2,k,i})$, $k \in \{0,1,\ldots,p\}, \  i  \in \{1,\ldots,d_k\}$ associated with ${\mathcal L}_{1,k}  \times {\mathcal L}_{2,k} \subset G_{(k)}$. 
Let $\mathcal H_k$ in \eqref{eq:def_setHk} be given in  terms  of these $v_{k,i}$,  and let   $x_\mathrm{ss} \in \real^n$ satisfy \eqref{eq:x_ss_condition}. Then there exist feedback matrices $F_q$ that solve Problem \ref{prob:nonovershoot},  with the set of initial states $\mathcal X_0$ given  by
\begin{align}\label{eq:def_setX}
\mathcal X_0 := \big\{ x_0 \in \real^n : \pi_k(x_0-x_\mathrm{ss}) \in {\mathcal H_k} \qquad\text{ for all $k \in \{0,1, \ldots, p\}$}  \big\}.
\end{align}
\end{theorem}%
\begin{proof}
For each $v_{k,i}$ and associated $\lambda_{q,k,i}$ we obtain $w_{q,k,i}$ from \eqref{MooreM2} and construct $V, W_q$ as in \eqref{MooreVW}.
Then the control law~\eqref{uqlaw} with $F_q$ obtained from \eqref{MooreF} simultaneously rectifies the subsystems $\Sigma_q$.
As $\mathcal L_{q,k}\subset\R^-$, the closed-loop system $\Sigma_\mathrm{CL}$ is GUAS, by  Theorem \ref{thm:GASAS}.

Consider $\xi_0 = x_0 - x_\mathrm{ss}\in\mathcal X_0$ and consider the closed-loop homogeneous switched system $\Sigma_{\xi}$ in \eqref{eq:sys_xi}.
Let  $\{t_j:  j \in \nat \} \subset [0, \infty)$ be the sequence  of switching instants, i.e. points of discontinuity of $\sigma$.
Consider the switching instant $t_j$ with $\sigma(t_j)=q$.
The feedback matrices $F_q$ together with subsystem \eqref{eq:sys_q} and partitioning $(d_0,d_1,\ldots,d_p)$ satisfy the assumptions of Theorem~\ref{thm:ea_LTI}.
Thus the outputs of the closed-loop system $\Sigma_\mathrm{\xi}$ in \eqref{eq:sys_xi} for $t\in[0,t_{j+1}-t_j)$ take the form 
\begin{align*}
e(t_j+t)& = \begin{bmatrix}      
	\alpha_{1,1} e^{\lambda_{q,1,1}t} + \dots  + \alpha_{1,d_1} e^{\lambda_{q,1,d_1} t} \\
	\vdots\     \\
	\alpha_{p,1} e^{\lambda_{q,p,1}t} + \dots  + \alpha_{p,d_p} e^{\lambda_{q,p,d_p}t} 
\end{bmatrix},
\end{align*}
where $\alpha_{k,i}$, $k\in\{1,\ldots,p\}$, $i\in\{1,\ldots,d_k\}$, are the last $n-d_0$ entries of $V^{-1}\xi(t_j)$.
Since $\alpha_{k,i}$, $i\in\{1,\ldots, d_k\}$, satisfy the conditions  in Lemma~\ref{lemA1} and \ref{lemA2} for $d_k=2$ and $d_k=3$, respectively, we conclude that $\pi(\xi_0) \in \mathcal H_k$ for each $k$,  and  
the output $e(t)$ is non-overshooting for $t\in[t_j,t_{j+1})$.

Note the eigenstructure of all $\Sigma_q$ is rectified and the output matrix $C$ is shared by all subsystems. 
Therefore we obtain the same $\alpha_{k,i}$, $k\in\{1\ldots, p\}$, $i\in\{1,\ldots,d_k\}$, for all subsystem $q\in\mathcal I$. 
Furthermore also $\xi(t)\in\mathcal X_0$ for $t\in[t_j,t_{j+1})$ which ensures that the initial condition $\xi(t_{j+1})\in\mathcal X_0$ guarantees non-overshooting of $e(t)$ for the subsequent switching interval.
Since $\Sigma_\mathrm{\xi}$ is GUAS $e(t)\rightarrow 0$.
Accordingly, the set-point tracking $y$ of $\Sigma_\mathrm{CL}$ in \eqref{eq:sysCL} is non-overshooting.
\end{proof}

\begin{remark}
We utilise the Lemmas \ref{lemA1} and \ref{lemA2} from \cite{SchN2010} to provide the Conditions I--III. for the absence of a sign change for a sum of two or three exponential functions.
Note that only Conditions~I and II in Lemma \ref{lemA2} are necessary and sufficient whereas Condition III is only sufficient.
Condition III may  be replaced by a pair of weaker  conditions that are together necessary and sufficient; we refer the reader to \cite[Lemma A.2]{SchN2010} for the details.
\end{remark}

\begin{remark}
Note that the control design in Theorem~\ref{thm:Final_nous} guarantees GUAS for every initial condition $x_0 \in \mathbb R^n$ and arbitrary switching signals.

As in the LTI case \cite{SchN2010}, $\mathcal X_0$ may not contain a neighbourhood of the equilibrium $x_\mathrm{ss}$.
In case the initial condition does not satisfy the condition on the parameters $\alpha_{k,i}$ for $d_k=2$ in \eqref{eq:def_setHk}, i.e. $x_0\notin\mathcal X_0$, there exist switching sequences leading to  an  overshooting output; in particular the constant switching signals $\sigma(t) \equiv 1 $  and  $\sigma(t) \equiv 2 $ will   yield  overshoot. 
Note however, there may exist many switching sequences $\sigma$ where no overshoot is observed for $x_0\notin\mathcal X_0$.
The same holds for initial conditions that violate Condition~I or II for $d_k=3$.
As Condition III in Lemma \ref{lemA2} is only sufficient, violation of this property does not even guarantee that there exists a switching sequence yielding an overshooting output.
\end{remark}

\begin{remark}
The rate of convergence of the output at any time $t>0$ is determined by the activated closed-loop poles from 
${\mathcal L}_q$ assigned to the 
eigenvectors in $\mathcal{R}^*_{q,(k)}$ for $k\neq 0$.
In particular, choosing identical eigenvalues for both closed-loop subsystems corresponding to the same output component results in a smooth output trajectory rendering the switching nature of the system dynamics invisible at the output in the sense that the outputs are continuously differentiable.
\end{remark}

The following results can be considered as a special case of Theorem \ref{thm:Final_nous} requiring stricter conditions on the system structure of the  constituent subsystems.
We show that the set-point tracking can be rendered globally monotonic for arbitrary initial conditions.

\begin{theorem} [Monotonic outputs]\label{thm:Final_monotonic}
Let the  switched system $\Sigma_{\rm OL}$ satisfy Assumption \ref{Ass1}.
Let the  $(p,n)$-partitioning with $d_{0}=n-p$, $d_{k}= 1$, $k\in\{1, \ldots, p\}$ be  feasible for $\Sigma_\mathrm{OL}$,  with  compatible  sets of desired  closed-loop  eigenvalues  $\mathcal L_{q,k} \subset \R^-$ be such  that  ${\mathcal L}_{1,k}  \times {\mathcal L}_{2,k} \subset G_{(k)}$,  and satisfy \eqref{eq:test_overallfeasibility}.
Then there exist feedback matrices $F_q$ that solve Problem \ref{prob:monotonic}.
\end{theorem}

\begin{proof}
With the definition of $d_{(0)}$ in \eqref{eq:def_d(k)} any choice $(\lambda_{1,0,i},\lambda_{2,0,i})\in G_{(0)}$ for distinct eigenvalues $\lambda_{1,0,i}\in\mathcal{L}_1, \lambda_{2,0,i}\in \mathcal{L}_2 $, yields linearly independent eigenvectors $v_{0,i}\in\im P_{(0)}(\lambda_{1,0,i},\lambda_{2,0,i})$, $i\in\{1,\ldots,n-p\}$.

Since Condition \eqref{eq:test_overallfeasibility} is satisfied we can choose the remaining $p$ eigenvectors $v_{k,1}\in \im P_{(k)}(\lambda_{1,k,1},\lambda_{2,k,1})$ with $\lambda_{q,k,1}\in\mathcal L_{q,k}$, for $k\!\in\!\{1,\ldots, p\}$ such that $\big\{v_{k,i}: k\!\in\!\{0,1,\dots,p\}, i\!\in\!\{1,\dots, d_k\}\big\}$ is a set of linearly independent vectors by Theorem \ref{thm:Switched_Prob2_solved}.
For each $v_{k,i}$ and associated $\lambda_{q,k,i}$ we obtain $w_{q,k,i}$ from \eqref{MooreM2}.
Construct $V, W_q$ as in \eqref{MooreVW}.
Then the control law \eqref{uqlaw} with $F_q$ obtained from \eqref{MooreF} simultaneously rectifies the subsystems $\Sigma_q$.
Since $\mathcal L_q\in\R^-$ we have GUAS for $\Sigma_\mathrm{CL}$ with Theorem \ref{thm:GASAS}.

In each switching instant the outputs of $\Sigma_\xi$ take the form \eqref{eq:output_ea_LTI}, where only the first coefficient $\alpha_{k,1}$, $k\in\{1,\ldots,p\}$ is non-zero.
Hence each output-component for any subsystem is monotonically decreasing from any initial condition.
Similar reasoning as in the proof of Theorem~\ref{thm:Final_nous} yields globally monotonic step-reference tracking of $\Sigma_\mathrm{CL}$ for arbitrary switching.
\end{proof}%

Theorem \ref{thm:Final_monotonic} can be considered as  a special case of Theorem \ref{thm:Final_nous} in the sense that a fixed partitioning is applied for the eigenstructure assignment.
Such partitioning together with the assignment of negative real eigenvalues ensures monotonic tracking of the output.
Furthermore, the monotonic tracking is guaranteed for any initial condition $x_0\in\R^n$.

As no test on the initial states is necessary the $d_0$ eigenvalues in $\mathcal L_{q,0}$ do not need to be sorted as in \eqref{eq:indexed_lambda_q}.
Therefore we may calculate the intersection \eqref{eq:intersection} for arbitrary pairs of the sets $\mathcal L_{1,0}$ and $\mathcal L_{2,0}$ which may increase the choice of suitable eigenvectors and thus provide more freedom in the design.

For the monotonic design the requirements regarding the intersections \eqref{eq:sum_of_Pks} are more restrictive than for the non-overshooting design, e.g. we require $d_{(0)}\geq n-p$.
Thus a given switched system \eqref{eq:sysOL} may be infeasible for the eigenstructure assignment for a monotonic response but a non-overshooting design may succeed.
The non-overshooting design offers more flexibility for the assignment of modes to the respective outputs. 
Here the size $m_{(0)}=m-p$ is irrelevant since no mode needs to be assigned to the kernel of $C$.
Moreover we observe that the intersections~\eqref{eq:intersection} for $k\in\{1,\ldots,p\}$
are typically larger than for $k=0$.
Consequently, there is a significant class of systems that can be considered for the non-overshooting set-point tracking but not for the monotonic set-point tracking.

Obviously the non-overshooting design offers more flexibility in choosing the partitioning $(d_0,d_1,\ldots,d_k)$ for which we only require that they sum to $n$ and $0\leq d_k\leq 3$.
Therefore monotonic and non-overshooting tracking can be combined in the same design to different outputs, e.g.
choosing $d_k=1$, monotonic tracking is assigned to the output $y_k$, regardless of the initial condition.

We  propose the following algorithms  to  obtain the feedback matrices for this  task.
The first algorithm provides the analysis  of the switched system and a feasible choice of partitioning for the design, whereas the second yields the feedback matrices for the control law and domain of initial states if needed.
\begin{alg}[Analysis]
\label{alg:analysis}\mbox{}

\noindent
\emph{(Inputs):} Systems  $\Sigma_q$, $q\in\mathcal I$ in \eqref{eq:sys_q} satisfying  Assumption~\ref{Ass1}.

\noindent
\emph{(Outputs):} A set $\mathcal D$ of feasible $(p,n)$-partitionings for  $\Sigma_\mathrm{OL}$. 

\begin{enumerate}
\item \label{alg:step_compute_RNMTP}
For each  $k\in\{0,1,\ldots, p\}$ and $q\in\mathcal I$,  compute the Rosenbrock matrices $R_{q,(k)}(\lambda)$,
matrix kernels $N_{q,(k)}$ and $M_{q,(k)}$,  transformation matrices $T_{(k)}(\lambda_1,\lambda_2)$,  dense sets $G_{(k)}$   and rational matrix functions 
$P_{(k)}(\lambda_1,\lambda_2)$,  as in Section \ref{sec:EA_switched}. 
\item \label{alg:step_compute_d(k)}
Compute $d_{(k)}$ as in  \eqref{eq:dkviarank}.\\ 
If $d_{(0)}+\sum_{k=1}^p \min\{3,d_{(k)}\}<n$: stop.
\item[(3a)] \textbf{\textit{For non-overshooting design:}} 
Collect in $\mathcal D$ all feasible partitionings $(d_0, d_1,\dots, d_p)$, satisfying   \eqref{eq:feas_swsys};
\item[(3b)] \textbf{\textit{For monotonic design:}} 
Collect in $\mathcal D$ all feasible $(p,n)$-partitionings $(n-p,1,\ldots,1)$ satisfying   \eqref{eq:feas_swsys};
\end{enumerate}
\end{alg}

\begin{remark}
If the condition in Step 2 fails for the transformation matrix $T_{(k)}(\lambda_1,\lambda_2)$ obtained in Step 1, then  the algorithm stops since no feasible partitioning exists for this representation.
However, it may be possible to find a different representation of the reduced row-echelon form at points $(\lambda_1^\star,\lambda_2^\star)$ with corresponding  $T_{(k)}^\star(\lambda_1^\star,\lambda_2^\star)$, (see Remark \ref{rem:defect}), for which 
\begin{align*}
d_{(0)}+\sum_{k=1}^p \min\{3,d_{(k)}\}  \geq n        
\end{align*}
and then a feasible partitioning vector may  be  found.
\end{remark}

\begin{alg}[Synthesis]
\label{alg:synthesis}\mbox{}

\emph{(Inputs):}
Systems  $\Sigma_q$, $q\in\mathcal I$ in \eqref{eq:sys_q} satisfying  Assumption~\ref{Ass1}.
A  feasible $(p,n)$-partitioning $(d_0, d_1,\dots, d_p)$ for  $\Sigma_\mathrm{OL}$. Sets of compatible desired  closed-loop  eigenvalues  $\mathcal L_{q,k} \subset \R^-$  such that $\mathcal{L}_{1,k}\times\mathcal{L}_{2,k} \in G_{(k)}$.

\emph{(Outputs):} Feedback matrices $F_q$ solving Problem \ref{Prob:estruct2}. A domain of initial states $\mathcal X_0 \subset \real^n $.  
\begin{enumerate}
\setcounter{enumi}{3}
\item Compute $\mathcal{P}_{(k)} ( \mathcal L_{1,k}\!\times\! \mathcal L_{2,k}  )$ in \eqref{eq:sum_of_Pks}.\\
\emph{If:} Condition \eqref{eq:test_overallfeasibility} in Theorem \ref{thm:Switched_Prob2_solved} holds continue,\\ 
\emph{Else:} choose a different feasible partitioning and/or eigenvalues $\mathcal{L}_{1,k}\times\mathcal{L}_{2,k} \in G_{(k)}$.\\
\emph{Stop:} if non of the partitionings  in $\mathcal D$ satisfies \eqref{eq:test_overallfeasibility}.
\item For $k\in\{0,1,\ldots,p\}$ and $i\in\{1,\ldots,d_k\}$ select $n$ linearly  independent  vectors
\begin{align*} 
	v_{k,i}   \in \im   P_{(k)}(\lambda_{1,k,i},\lambda_{2,k,i}). 
\end{align*}
\item \label{alg:step_compute_Fq}
For each $v_{k,i}$  and  $\lambda_{q,k,i}$\,, $q \in \mathcal I$, $k \in \{0,1,\ldots,p\}$, $i  \in \{1,\ldots, d_k\}$, solve \eqref{MooreM2}  for $w_{q,k,i}$ to obtain matrices $V$,  $W_q$ and $F_q$  from  \eqref{MooreF}-\eqref{MooreVW}, ensuring that the indexing of the columns of  $V$  and $W_q$ correspond to the partitioning $(d_0, d_1,\dots, d_p)$ and match those of the $\mathcal L_{q,k}$. 
\item[7a)] \hypertarget{alg:step_init_cond_set}{} \textbf{For the non-overshoot design:} compute the set of initial states $\mathcal X_0$ in \eqref{eq:def_setX} using \eqref{eq:def_setHk}.
\item[7b)] \textbf{For the monotonic design:}  $\mathcal  X_0  = \R^n$. 
\end{enumerate}
\end{alg}

\section{Examples}\label{sec:example}
In this section we illustrate the design procedure and design options by two numerical examples adapted from \cite{HonSWR24}.
The first examples allows for non-overshooting design only, whereas for the second examples monotonic outputs can be achieved.

\subsection{Non-overshooting design} \label{sec:ex_nonovershoot}
Consider the subsystems $\Sigma_q$ with
\setlength{\arraycolsep}{1.5pt}
{\small
\begin{align*}
A_1\!\!&=\!\!\begin{bmatrix*}[c]
	0 & 1 & 0 & 0 & 0 & 0 & 0\\ 0 & -1 & 0 & 0 & 0 & 0 & 0\\ 0 & 0 & 0 & 0 & 0 & 0 & 0\\ -4 & 0 & 0 & -4 & 1 & -1 & 0\\ 2 & 0 & 0 & -4 & 0 & 0 & 0\\ 0 & 0 & -1 & 0 & 0 & 0 & 0\\ 0 & 0 & 0 & 0 & 0 & 0 & 0 
\end{bmatrix*}\!\!,    \
A_2\!\!=\!\!\begin{bmatrix*}[c]
	-5 & 0 & 2 & 0 & 0 & 0 & 0\\ 0 & 0 & 0 & 0 & 0 & 3 & 0\\ 0 & 4 & 0 & 0 & 0 & 0 & 0\\ 0 & 0 & 0 & 0 & 0 & -3 & 0\\ 4 & 0 & 0 & -5 & 0 & 3 & 0\\ 0 & 0 & -3 & 0 & 0 & 0 & 0\\ 0 & 0 & 0 & 0 & 0 & -2 & 0 
\end{bmatrix*}\!\!,
B_1\!\!&=\!\!\begin{bmatrix}
	3 & 0 & -3 & 0 & 0\\ 0 & 1 & -1 & 4 & 0\\ -1 & 0 & 0 & -2 & 1\\ 0 & 0 & 0 & 0 & 5\\ 0 & 5 & 3 & 0 & 0\\ 4 & 0 & 0 & 0 & 0\\ -4 & 0 & 0 & -4 & 3
\end{bmatrix}\!\!,\   
B_2\!\!=\!\!\begin{bmatrix}
	0 & 0 & 5 & 0 & -5\\ 0 & 2 & 2 & 0 & 0\\ 0 & 0 & 0 & 0 & -2\\ 1 & -4 & 0 & 0 & 0\\ 0 & 0 & 0 & 0 & 0\\ 0 & 0 & 0 & 0 & 0\\ 0 & 0 & 0 & -1 & 0 
\end{bmatrix}\!\!, \\ 
C\!\!&=\!\!\begin{bmatrix} 1 & 1 & 0 & 0 & 0 & 0 & 0\\ 0 & 0 & 0 & 0 & 0 & 0 & 1\\ 0 & 1 & 0 & 0 & 0 & 0 & 0  \end{bmatrix}\!\!,
\end{align*}%
}%
and the set-point $r=\begin{bmatrix}1&-6& 10\end{bmatrix}^\top$ to be tracked.

With \eqref{eq:x_ss_condition} we obtain the joint steady state
$x_\mathrm{ss}=\begin{bmatrix} -9& 10&0 &14.67  &59.11 & 36.44&-6 \end{bmatrix}^\top$, as well as steady-state control for each subsystem
\begin{align*}
u_{1,\mathrm{ss}}&=\begin{bmatrix}0 &13.33 &3.33 &0  &0 \end{bmatrix}^\top,\\
u_{2
,\mathrm{ss}}&=\begin{bmatrix} -153.33& -65.67& 11& -72.89 & 20\end{bmatrix}^\top.
\end{align*}%

For these subsystems, we  have $n = 7$, $m = 5$ and  $p=3$,  so Assumption  \ref{Ass1} on the system dimensions  
is satisfied.

Evaluating Step~\ref{alg:step_compute_RNMTP} of Algorithm \ref{alg:synthesis} we note that the intersection space  for $k=0$
is
\begin{align*}
\im P_{(0)} (\lambda_1,\lambda_2)=0.
\end{align*}
Therefore $d_{(0)}=0$ and thus no mode can be hidden from all outputs.
The other matrices $P_{(k)}(\lambda_1,\lambda_2)$ and the sets $G_{(k)}$ for $k\in\{1,\ldots,3\}$ are obtained as:
{\small
\begin{align*}
G_{(1)}=&\big(\mathbb R\setminus \sigma(A_1)\times \mathbb R\setminus \sigma(A_2)\big)\setminus 
\big\{(\lambda_1,\lambda_2)|(74\lambda_1 \!-\!36\lambda_2 \!+\!13\lambda_1 \lambda_2 \!+\!4\lambda_1 ^2\lambda_2 \!+\!6\lambda_1 ^2+9=0)\big\} \\
G_{(2)}=&\big(\mathbb R\setminus \sigma(A_1)\times \mathbb R\setminus \sigma(A_2)\big)\setminus  
\big\{(\lambda_1,\lambda_2)|(12\lambda_2 -5\lambda_1 +2\lambda_1\lambda_2 +10=0)\big\} \\
G_{(3)}=&\big(\mathbb R\setminus \sigma(A_1)\times \mathbb R\setminus \sigma(A_2)\big)\setminus
\big\{(\lambda_1,\lambda_2)|(23\lambda_1^2\lambda_2 ^3+372\lambda_1 ^2\lambda_2 ^2-582\lambda_1 ^2\lambda_2 -342\lambda_1 ^2\\ 
&\phantom{xxxxxxxx}+109\lambda_1\lambda_2^3 +2137\lambda_1\lambda_2 ^2+2382\lambda_1\lambda_2 +702\lambda_1
+20\lambda_2 ^3-11\lambda_2^2-174\lambda_2 +432=0)\big\}.
\end{align*}
}

Using \eqref{eq:dkviarank} as in Step \ref{alg:step_compute_d(k)} it is readily verified that  $d_{(k)}=5$ for all $k\in\{1,\ldots,3\}$ and therefore five linearly independent eigenvectors can be obtained for each intersection $\im N_{1,(k)}(\lambda_1)\cap \im N_{2,(k)}(\lambda_2)$.
Therefore the $(p,n)-$partitioning $(d_0,d_1,d_2,d_3)=(0,3,3,1)$ is a feasible choice for $\Sigma_{\rm{OL}}$ according to Definition \ref{def:feas_swsys}.
Further we choose the indexed eigenvalue sets 
\begin{align*}
\mathcal{L}_{1,1}\times\mathcal{L}_{2,1} &=\{-5,-4,-3\} \times \{-3,-2,-1\} \in G_{(1)}\\
\mathcal{L}_{1,2}\times\mathcal{L}_{2,2} &=\{-8,-7,-6\} \times \{-7,-6,-4\} \in G_{(2)}\\
\mathcal{L}_{1,3}\times\mathcal{L}_{2,3} &=\{-0.5\} \times \{-8\} \in G_{(3)}.
\end{align*}%
It is readily verified that Condition \eqref{eq:test_overallfeasibility} holds. 
Thus we can choose $n$ linearly independent eigenvectors from the intersections obtained from the above sets of eigenvalues.
Accordingly we choose three eigenvectors of the output-nulling subspace $v_{1,i}  \in \im     P_{(1)}(\lambda_{1,1,i},\lambda_{2,1,i})$, $i\in\{1,2,3\}$, three from the output-nulling subspace $v_{2,i}  \in \im     P_{(2)}(\lambda_{1,2,i},\lambda_{2,2,i})$, $i\in\{1,2,3\}$, and finally $v_{3,1}\in\im     P_{(3)}(\lambda_{1,3,1},\lambda_{2,3,1})$.

Using Step \ref{alg:step_compute_Fq} of Algorithm \ref{alg:synthesis}, we obtain the feedback matrices for each subsystem as
\setlength{\arraycolsep}{1.3pt}
{\footnotesize 
\begin{eqnarray*}
	F_1\!=\!\begin{bmatrix*}[r] 
		0.08     &    -0.49   &      -3.07 &         0.09    &     -0.03   &       0.13       &   0.65\\
		-1.84   & -1.70  &  -7.48    &  1.44   &   -1.16       &   6.82       &  -1.06\\
		6.47 &   -0.72  & -49.38   &   -1.24   &  1.02  &  1.92   &   18.59\\
		2.08   &  0.37   &  -10.48     & -0.67   & 0.54  &   -1.23  &   4.91\\
		14.40  &  -6.69  &  -129.51 &  -3.65  & 3.99   &  -2.55  & 46.13
\end{bmatrix*}\end{eqnarray*}}%
and
{\footnotesize
\begin{eqnarray*}
	F_2\!=\!\begin{bmatrix*}[r]  
		102.82   &  -67.22   & -867.73   &-38.18 &        34.90   &  -63.96  &  333.11\\
		7.99  &   -10.17  &  -77.42   &-2.50 &  2.41    &  -2.80  & 28.99\\
		-7.99 &   6.17   &  77.42  &2.50 &     -2.41    &   1.30  &  -28.99\\
		-37.71   &  15.54  &  330.71  &11.51   &     -11.83  &  13.69&  -122.41\\
		-5.55 &  3.93  &   50.03  &    1.70  &     -1.79   &  2.38 &  -18.23
	\end{bmatrix*}\!.
\end{eqnarray*}}%
By Theorem \ref{thm:Final_nous}, the  outputs of $\Sigma_{\rm {CL}}$ will  track this reference without overshooting if $x_0-x_\mathrm{ss}\in\mathcal X_0$.
The set of initial conditions $\mathcal X_0$ is obtained as in \eqref{eq:def_setX} evaluating \eqref{eq:def_setHk} as in Step \hyperlink{alg:step_init_cond_set}{7a}.

Note for the evaluation of the conditions on $\alpha_{k,i}$ in \eqref{eq:def_setHk} the eigenvalues assigned for each output have to be sorted in ascending order \eqref{eq:indexed_lambda_q} and paired accordingly, e.g. the eigenvalue $\lambda_{1,1,1}=-5$ has to be paired with the eigenvalue $\lambda_{2,1,1}=-3$, etc.
To the third output only one mode is assigned such that this output tracks its reference monotonically for arbitrary initial state.
Note, that also for $k=0$ we have no constraints on the initial state.
Therefore the sets $\mathcal L_{q,0}$, if they exist, are not indexed such that the eigenvalues may be paired arbitrarily, providing additional degrees of freedom for the design see Example \ref{sec:ex_monotonic}.

Of course we can also choose a different partitioning that satisfies $\sum d_k=n$ and $d_k\leq \min\{3,d_{(k)}\}$.
This might be necessary, e.g. in case Condition \eqref{eq:feas_swsys} fails for some initial choice of partitioning.
For instance the partitioning $(d_0,d_1,d_2,d_3)=(0,3,2,2)$ is also feasible and satisfies Condition \eqref{eq:feas_swsys}.
Such design would lead to non-overshooting, but not necessarily monotonic, step-reference tracking in all three outputs.
Since $d_{(0)}=0$ and we have to allocate seven modes to three outputs with maximal three modes to each output.
Thus we can only choose one output for a monotonic design.

\begin{figure}[tb]
\centering
\psfrag{y1}[][][0.6]{\Large$y_1$}
\psfrag{y2}[][][0.6]{\Large$y_2$}
\psfrag{y3}[][][0.6]{\Large$y_3$}
\psfrag{y}[][][0.6]{\Large$y$}
\psfrag{r1}[][][0.6]{\Large$r_1$}
\psfrag{r2}[][][0.6]{\Large$r_2$}
\psfrag{r3}[][][0.6]{\Large$r_3$}
\psfrag{t}[][][0.6]{\Large$t$}
\psfrag{s(t)}[][][0.6]{\Large$\sigma(t)$}
\includegraphics[width=.7\linewidth]{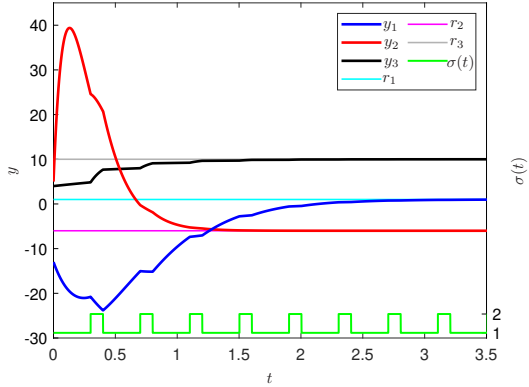}
\caption{Step reference tracking for the rectified eigenstructure design with non-overshooting outputs.}
\label{fig:LessCoutput}
\end{figure}

Figure \ref{fig:LessCoutput} shows a simulation of the closed-loop switched system for the partitioning $(d_0,d_1,d_2,d_3)=(0,3,3,1)$ above,
the initial state $x_0=\begin{bmatrix}-17&4&0& 2&-7&-3&5\end{bmatrix}^\top$, and the periodic switching signal that activates $\Sigma_1$ for 0.3s followed by  $\Sigma_2$ for the subsequent 0.1s.
It is readily verified that the parameters $\alpha_{k,i}$ obtained from $V^{-1}(x_0-x_\mathrm{ss})$ satisfy the conditions in \eqref{eq:def_setHk}.
We observe that $y_1$ and $y_2$ are non-overshooting, whereas $y_3$ exhibits even monotonic behaviour.

\subsection{Monotonic design} \label{sec:ex_monotonic}

For a second example we consider the dynamics $A_1,A_2$ and input matrices $B_1,B_2$ of Example \ref{sec:ex_nonovershoot} but remove one output and consider the output matrix
\begin{align*}C=\begin{bmatrix} 1 & 1 & 0 & 0 & 0 & 0 & 0\\0 & 1 & 0 & 0 & 0 & 0 & 0 \end{bmatrix}  
\end{align*}
instead.
Since the system matrices $A_q$, $B_q$ remain unchanged, but we  only consider a two component reference vector $r=\begin{bmatrix}
1&-6
\end{bmatrix}^\top$ we obtain the steady-state $x_\mathrm{ss}=\begin{bmatrix}
7&-6&0&-8&-26.67&-22.67&0
\end{bmatrix}^\top$ and steady-state control $u_{1,\mathrm{ss}}=\begin{bmatrix}
0&-8&-2&0&0
\end{bmatrix}^\top$ and $u_{2,\mathrm{ss}}=\begin{bmatrix}
88&39&-5&45.33&-12
\end{bmatrix}^\top$ .

For this example we obtain $d_{(0)}=5$ and thus a monotonic design might succeed.
We verify that the partitioning $(d_0,d_1,d_2)=(5,1,1)$ is feasible for $\Sigma_{\mathrm{OL}}$.

After calculating feasible sets of eigenvalues $G_{(k)}$, which are not displayed due to space restrictions, we choose the following sets of eigenvalues
\begin{align*}
\mathcal{L}_{1,0}\!\times\!\mathcal{L}_{2,0} &=\{-7 ,   -3  ,  -4 ,   -5  ,  -6\} \times \{-2  ,  -1   , -3   , -4   , -6\} \in G_{(0)}\\
\mathcal{L}_{1,1}\!\times\!\mathcal{L}_{2,1} &=\{-0.5\} \!\times\! \{-7\} \in G_{(1)}\\
\mathcal{L}_{1,2}\!\times\!\mathcal{L}_{2,2} &=\{-8\} \!\times\! \{-8\} \in G_{(2)}.
\end{align*}
Note that the assigned closed-loop eigenvalues are the same as in the first example.
However, the sets need not to be indexed such that arbitrary pairing is possible.
While this offers more flexibility for performance design of the switched system, it also gives more options in case the intersections \eqref{eq:sum_of_Pks} are not rich enough to span the whole state-space.
A different pairing of the eigenvalues may lead to larger intersections such that Condition \eqref{eq:test_overallfeasibility} is satisfied.

\begin{figure}[tb]
\centering
\psfrag{y1}[][][0.6]{\Large$y_1$}
\psfrag{y2}[][][0.6]{\Large$y_2$}
\psfrag{y3}[][][0.6]{\Large$y_3$}
\psfrag{y}[][][0.6]{\Large$y$}
\psfrag{r1}[][][0.6]{\Large$r_1$}
\psfrag{r2}[][][0.6]{\Large$r_2$}
\psfrag{s(t)}[][][0.6]{\Large$\sigma(t)$}
\psfrag{t}[][][0.6]{\Large$t$}
\includegraphics[width=.7\linewidth]{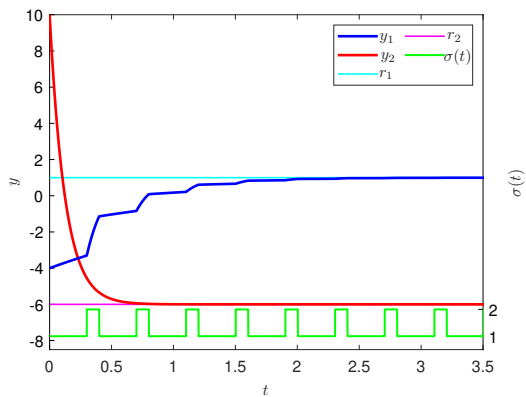}
\caption{Step reference tracking for the rectified eigenstructure design with monotonic outputs.}
\label{fig:MonotonicDesign}
\end{figure}

Note further that there is no restriction for the initial condition $x_0$, such that Algorithm \ref{alg:synthesis} terminates with Step~\ref{alg:step_compute_Fq}.
To highlight that we choose an initial condition 
that does not satisfy the conditions on $\alpha_{k,i}$ in \eqref{eq:def_setHk}:
$x_0=\begin{bmatrix}
-14&10&-1&2&52&35&-9
\end{bmatrix}^\top$.
The resulting response is shown in Figure \ref{fig:MonotonicDesign} with the same switching signal and reference as before.
We observe that both outputs are monotonic. 
Moreover $y_2$ is even a smooth function, i.e. the switching instances are invisible. 
This is achieved by assigning identical eigenvalues for the modes assigned to the second output.

\section{Conclusion}
We investigate a state-feedback design approach for switched linear MIMO systems.
By analysing eigenstructure properties of the constituent subsystems we obtain feasible partitionings that jointly assign modes to the respective outputs of the system.
We provide a structural condition for the feasibility of the combined problem of mode-assignment to the outputs and eigenvector rectification of the subsystems.
Our analysis provides a choice of partitionings, i. e. possibilities to assign certain properties to the respective outputs as well as sets of eigenvectors and eigenvalues to be assigned.
The proposed design procedure assigns selected eigenvectors and eigenvalues to the closed-loop subsystems.
This achieves eigenvector rectification for the constituent subsystems while assigning closed-loop poles in the left complex half-plane.
By that we obtain stability  for arbitrary switching as well as non-overshooting or monotonic output tracking for the specified outputs.
While this contribution is focussed  on the design for two subsystems the methodology may be applicable to a larger number of subsystems $q\geq 3$. However the complexity of the expressions increase rapidly and conditions may turn out so strict that no feasible design may exist.

\section{Appendix}

\begin{lemma}
\label{lemA1} \cite[Lemma A.1]{SchN2010}
Let $\lambda_1 < \lambda_2 < 0$ and let 
\begin{align*}
	f(t) = \alpha_1\, e^{\lambda_1 t} + \alpha_2\, e^{\lambda_2  t}.
\end{align*}
Then $f(t)$  does not change sign for all $t \geq 0
$ if and only if
$(\alpha_{1} + \alpha_2)\alpha_2 > 0 $.
\end{lemma}

\begin{lemma} 
\label{lemA2} \cite[Lemma A.2]{SchN2010}
Let $\lambda_1 < \lambda_2 < \lambda_3 < 0$, and define
\begin{align*} \label{fct}
	f = \alpha_1\, e^{\lambda_1 t} + \alpha_2\, e^{\lambda_2 t} +
	\alpha_3\, e^{\lambda_3 t}.
\end{align*}
Then $f(t)$ does  not  change sign for any $t > 0$ if none of 
the following  conditions hold:
\bn
\item[I.] $\alpha_1 \, \alpha_2 >0$, $\alpha_1 \,\alpha_3 <0$, and $|\alpha_{1} + \alpha_2 | >
|\alpha_3|$;
\item[II.] $\alpha_2 \, \alpha_3 >0$,
$\alpha_1 \, \alpha_2 <0$, and $|\alpha_{1} | > |\alpha_2 + \alpha_3|$;
\item[III.] $(\alpha_2+\alpha_3)\alpha_3<0$.
\en
\end{lemma}
The last condition in Lemma \ref{lemA2} is replaced by an equivalent expression to the one given in \cite[Lemma A.2]{SchN2010}.

\begin{lemma}[Rad\'{o}'s Theorem \cite{Rado42}]\label{lem:rado}
Let $\mathcal P_1,\ldots, \mathcal P_s$ be sets of an Euclidean space. 
There exist elements $p_i \in \mathcal P_i$ for all $i \in\{1,\ldots, s\}$ such that $\{p_1,\ldots, p_s \}$ is a set of linearly independent vectors if and only if given $k$ numbers $\nu_1 ,\ldots , \nu_k$ such that $1 \leq \nu_1 < \nu_2 <\ldots < \nu_k \leq s$ for all $k \in\{1,\ldots , s\}$, the union $\mathcal P_{\nu_1}\cup \mathcal P_{\nu_2}\cup \ldots \cup \mathcal P_{\nu_k}$ contains $k$ linearly independent elements.
\end{lemma}

\begin{prop}\label{prop:rado}
Let  the integer-valued tuple  $(d_0, d_1, \dots, d_p) $  be such that 
$\sum_{k = 0}^p \ d_k =  n$.
Let $\mathcal P_0,\mathcal P_1,\ldots,\mathcal P_p$ be subspaces of $\R^n$, such that 
$\dim \mathcal P_k=   d_k $.
There exists a set of linearly independent vectors $\{p_{k,i}  : k \in \{0,1\ldots,p\} \mbox{ and } i \in \{1\ldots,d_k\}\}$
such that each $p_{k,i}  \in  \mathcal P_k$ if and only if 
\begin{equation*}
	\dim \bigl(\mathcal P_0+\sum_{k \in S} \mathcal P_k \bigr) \ge d_0 + \sum_{k \in S} d_k  \quad \forall {S}\in2^{[1,p]}.
\end{equation*}
\end{prop}
\begin{proof}
Consider constants $\nu_i$ with $1 \leq \nu_1 < \ldots < \nu_k \leq s$.
Since $\sum \mathcal P_i=\Span \bigcup \mathcal P_i$, we have: $\mathcal P_{\nu_1}\cup  \ldots \cup \mathcal P_{\nu_k}$ contains $k$ linearly independent vectors if and only if $\mathcal P_{\nu_1}+  \ldots + \mathcal P_{\nu_k}$ contains $k$ linearly independent vectors. 
Thus $\dim (\mathcal P_{\nu_1}+  \ldots + \mathcal P_{\nu_k}) = k$.
With Lemma \ref{lem:rado} it follows that there exist $p_i\in\mathcal P_i$ such that $\{p_1,\ldots,p_s\}$ is linearly independent if and only if
\begin{align}\label{eq:proof_card}
	\forall S\in2^{[1,s]}: \dim \sum_{j\in S}\mathcal P_j\geq \mathrm{card}(S).
\end{align}
Consider the direct sum $\mathcal P_0 \oplus \mathcal P_k$, $k \in \{1\ldots,p\}$.
Let $\begin{bmatrix} I_{d_0}\\ 0 \end{bmatrix}$ be a basis of $\mathcal P_0$ and $\begin{bmatrix} P_{k,1}\\ P_{k,2} \end{bmatrix}$ a basis of $\mathcal P_k$, $k \in \{1\ldots,p\}$.
We can find a set of linearly independent vectors $\{p_{01}, \ldots, p_{0d_0} , p_1, \ldots , p_s \}$ such that $\Span \{p_{01}, \ldots, p_{0d_0}\} \subset \mathcal P_0$ and $p_j \in \mathcal P_j$ for all $j \in \{1, \ldots, s\}$ if and only if there exist $\tilde p_j\in\mathcal P_{j,2}$  such that $\{\tilde p_1, \ldots, \tilde p_s\}$ is linearly independent.
With \eqref{eq:proof_card} the assertion of the proposition follows.
\end{proof}

\bibliographystyle{plain}
\bibliography{../../sample,../../stability_tidy,../../switched_systems_applications,../../switched_systems_tracking_control}

\end{document}